\documentclass[acmsmall]{acmart}

\usepackage{booktabs} % For formal tables
\usepackage{flushend}
\usepackage[ruled]{algorithm2e} % For algorithms
\usepackage{multirow}
\usepackage{subfigure}
\usepackage{tcolorbox,threeparttablex,longtable,afterpage}
\usepackage{latexsym}
\usepackage{amsmath}
\usepackage{enumitem}
\usepackage{color}
\usepackage{xcolor,colortbl}
\usepackage{tcolorbox}
\newcommand{\tabincell}[2]{\begin{tabular}{@{}#1@{}}#2\end{tabular}}

\usepackage{tikz}
\newcommand*{\circled}[1]{\lower.7ex\hbox{\tikz\draw (0pt, 0pt)%
    circle (.5em) node {\makebox[1em][c]{\small #1}};}}

\newcommand{\finding}[1]{\begin{normalsize} \begin{tcolorbox}#1\end{tcolorbox} \end{normalsize}}

\AtBeginDocument{%
  \providecommand\BibTeX{{%
    \normalfont B\kern-0.5em{\scshape i\kern-0.25em b}\kern-0.8em\TeX}}}

\setcopyright{acmcopyright}
\acmJournal{TOSEM}
\acmYear{2023} \acmVolume{1} \acmNumber{1} \acmArticle{1} \acmMonth{1} \acmPrice{15.00}

\begin{document}

%%
%% The "title" command has an optional parameter,
%% allowing the author to define a "short title" to be used in page headers.
% \title{Emoji-Powered Sentiment Analysis and Emotion Mining in Software Engineering}
% \title{A Comprehensive Empirical Study of Bias Mitigation Methods for Software Fairness}
\title{A Comprehensive Empirical Study of Bias Mitigation Methods for Machine Learning Classifiers}
% (**) MH says: I propose this title, and in particular suggest removing the word "revisiting" which makes it sound very incremental and not novel. Also getting in the word empirical study makes it sound more like a software engineering paper, and adding comprehensive makes it more obvious what the primary claimed contribution is

%%
%% The "author" command and its associated commands are used to define
%% the authors and their affiliations.
%% Of note is the shared affiliation of the first two authors, and the
%% "authornote" and "authornotemark" commands
%% used to denote shared contribution to the research.

\author{Zhenpeng Chen}
\affiliation{%
  \institution{University College London}
  \city{London}
  \country{United Kingdom}
}

\author{Jie M. Zhang}
\affiliation{%
  \institution{King's College London}
  \city{London}
  \country{United Kingdom}
}

\author{Federica Sarro}
\affiliation{%
  \institution{University College London}
  \city{London}
  \country{United Kingdom}
}

\author{Mark Harman}
\affiliation{%
  \institution{University College London}
  \city{London}
  \country{United Kingdom}
}

%%
%% By default, the full list of authors will be used in the page
%% headers. Often, this list is too long, and will overlap
%% other information printed in the page headers. This command allows
%% the author to define a more concise list
%% of authors' names for this purpose.
\renewcommand{\shortauthors}{Z. Chen et al.}

%%
%% The abstract is a short summary of the work to be presented in the
%% article.
\begin{abstract}
Software bias is an increasingly important operational concern for software engineers.
We present a large-scale, comprehensive empirical study of 17 representative bias mitigation methods for Machine Learning (ML) classifiers, evaluated with 11 ML performance metrics (e.g., accuracy), 4 fairness metrics, and 20 types of fairness-performance trade-off assessment, applied to 8 widely-adopted software decision tasks. 
The empirical coverage is much more comprehensive, covering the largest numbers of bias mitigation methods, evaluation metrics, and fairness-performance trade-off measures compared to previous work on this important software property. 
We find that
(1) the bias mitigation methods significantly decrease ML performance in 53\% of the studied scenarios (ranging between 42\%$\sim$66\% according to different ML performance metrics);
(2) the bias mitigation methods significantly improve fairness measured by the 4 used metrics in 46\% of all the scenarios (ranging between 24\%$\sim$59\% according to different fairness metrics); 
(3) the bias mitigation methods even lead to decrease in both fairness and ML performance in 25\% of the scenarios;
(4) the effectiveness of the bias mitigation methods depends on tasks, models, the choice of protected attributes, and the set of metrics used to assess fairness and ML performance;
(5) there is no bias mitigation method that can achieve the best trade-off in all the scenarios.
The best method that we find outperforms other methods in 30\% of the scenarios. Researchers and practitioners need to choose the bias mitigation method best suited to their intended application scenario(s).
\end{abstract}

%%
%% The code below is generated by the tool at http://dl.acm.org/ccs.cfm.
%% Please copy and paste the code instead of the example below.
%%

%%
%% Keywords. The author(s) should pick words that accurately describe
%% the work being presented. Separate the keywords with commas.

%\thanks{Parts}
\authorsaddresses{Zhenpeng Chen, Federica Sarro, and Mark Harman are with the Department of Computer Science, University College London, London, United Kingdom. Emails: \{zp.chen, f.sarro, mark.harman\}@ucl.ac.uk. Jie M. Zhang is with the Department of Informatics, King's College London, London, United Kingdom. E-mail: jie.zhang@kcl.ac.uk. Zhenpeng Chen is the corresponding author.}
%%
%% This command processes the author and affiliation and title
%% information and builds the first part of the formatted document.

\maketitle

\section{Introduction}\label{intro}
Machine Learning (ML) software has made its way into a wide range of critical decision-making applications, such as hiring, criminal justice, credit risk prediction, and admissions \cite{csurMehrabiMSLG21}. 
There are several widely-known examples of software exhibiting unfair behaviour, relating to protected attributes such as gender \cite{faircase1,faircase3} and race \cite{faircase,faircase2}. 
Unfair software behaviour may result in unacceptable and unethical consequences that adversely affect users in minority and/or historically disadvantaged groups. 
Moreover, when software falls within legal or regulatory frameworks, unfair behaviour also incurs legal risks to software engineers.

The fairness issue has been studied for some time in Software Engineering (SE) research~\cite{afetal:re08}, pre-dating the recent upsurge in ML applications. 
The SE research community has increasingly focused on fairness since then. 
This increased focus is induced by increasing software systems' reliance on ML as a powerful generic technique to tackle complex decision and prediction problems, bringing with it the potential for unfairness as a result of software bias.\footnote{We use ``bias''  to refer to the opposite of ``fairness'' and treat ``unfairness'' and ``bias'' as synonyms.}
In recent years, the SE literature has witnessed a large number of results on software bias and fairness \cite{maxasefairness,sigsoftBrunM18,sigsoftHortZSH21,icseZhangH21,biswas2020machine,fairwaypaper,fairsmotepaper,sigsoftGalhotraBM17,UdeshiAC18,sigsoftAggarwalLNDS19,icseZhangW0D0WDD20,sigsoftBiswasR21,kbseChakrabortyPM20}. 

Software engineers tend to regard fairness as a non-functional property (typically most pertinent to ML software \cite{jieMLsurvey}, although also applicable more generally in SE \cite{afetal:re08}). In SE nomenclature, unfairness can be thought of ``fairness bugs'' \cite{Dabs220710223}, thereby motivating software engineers to seek techniques for fixing fairness bugs to reduce software bias (i.e., bias mitigation) \cite{sigsoftHortZSH21,maxasefairness,fairwaypaper,fairsmotepaper,biswas2020machine}.

With the emergence of various bias mitigation methods, SE researchers have started to evaluate and compare these methods \cite{sigsoftHortZSH21,fairwaypaper,fairsmotepaper,biswas2020machine,sigsoftBiswasR21}.
Such empirical evaluations enable the development of scientific knowledge about how useful different bias mitigation methods are in different application scenarios. 
There is a widespread belief in the research community that bias mitigation methods often improve fairness at the cost of ML performance (e.g., accuracy), known as ``\emph{fairness-performance trade-off}'' \cite{berk2021fairness,nipsWickpT19,sigsoftChenZSH22}. An effective bias mitigation method should improve fairness without decreasing ML performance too much, i.e., achieving a good trade-off between fairness and ML performance. Therefore, to evaluate the effectiveness of a bias mitigation method, researchers tend to not only measure the fairness improved by it, but also consider its effect on ML performance. 

However, existing evaluations of bias mitigation methods have yet to achieve full coverage and completeness to give a comprehensive picture.
In particular, researchers often use one or two metrics to measure the effects of existing bias mitigation methods on ML performance, overlooking other metrics that are widely used in industry and academia.
For example, Zhang and Harman \cite{icseZhangH21} and Hort et al. \cite{sigsoftHortZSH21} measure ML performance in terms of only accuracy; Chakraborty et al. \cite{fairwaypaper} use recall on favorable and unfavorable classes as only ML performance metrics; Biswas and Rajan~\cite{biswas2020machine,sigsoftBiswasR21} measure only accuracy as well as F1-score on the favorable class.

Nevertheless, different situations require different metrics.
For example, in an online advertising task targeting users with lower income, software engineers would care about precision for this category of users. It is because, with a fixed budget,  advertisers can place advertisements for a fixed number of users to view or click \cite{mangani2004online}. Among these users, advertisers want as many of them as possible to be low-income (i.e., a high precision of low-income users).
However, existing work does not consider the corresponding metric (i.e., precision on the unfavorable class) when evaluating bias mitigation methods on benchmark tasks such as the income prediction task \cite{icseZhangH21,sigsoftHortZSH21,fairsmotepaper,fairwaypaper}. Therefore, based on the results of existing work, software engineers will have no findings on which to base their decisions for such scenarios. 

This gap motivates us to evaluate existing bias mitigation methods using comprehensive metrics. 
Since different metrics measure the functional or non-functional properties of ML software from different aspects, we believe that the results of this study can provide insightful implications for real-world applications as well as a foundational baseline for further research and follow-on studies.

We present a comprehensive study, evaluating 17 representative bias mitigation methods in 8 widely-adopted benchmark tasks with 11 ML performance metrics, 4 fairness metrics, and 20 fairness-performance trade-off measures. 
Our study covers the largest number of bias mitigation methods, evaluation metrics, and fairness-performance trade-off measures in software fairness literature. 
Specifically, our study yields the following implications for fairness research and practice:

\begin {enumerate}
\item The values of all ML performance metrics that we use decrease in a large proportion (42\%$\sim$66\%) of scenarios after applying bias mitigation methods. Therefore, it is important for researchers to use a comprehensive set of task-relevant ML performance metrics in their evaluations, to be sure to capture any decrease in ML performance caused by bias mitigation methods.  

\item Existing bias mitigation methods improve fairness measured by the used metrics in 46\% of the scenarios that we study, and even lead to decrease in both fairness and ML performance in 25\% of the scenarios. Therefore, a community effort is required to bring software fairness improvement to a level where it becomes more effective and usable in practice.

\item The effectiveness of existing bias mitigation methods is affected by tasks, models, the choice of protected attributes, and the set of metrics used to measure fairness and ML performance; there is no bias mitigation method that can achieve the best trade-off in all the scenarios. Therefore, researchers and practitioners need to choose the suitable method according to their intended application scenario(s). Our results provide empirical guidance for such choices.

\item We have made publicly available all the scripts and data used in this study \cite{githublink} to allow for future replication and extension of our work.
\end{enumerate}

The rest of this paper is organized as follows. Section \ref{background} describes the background knowledge and motivation of this study. Section \ref{method} presents our research questions and methodology. Section \ref{results} reports and analyzes the results. Section \ref{threat} discusses threats to the validity. Section \ref{related_work} summarizes the related work, followed by concluding remarks in Section \ref{conclusion}.

\section{Preliminaries}\label{background}
This section provides background knowledge on software fairness and motivates this study.

\subsection{Definition of Fairness}
There are two primary types of fairness that researchers pursue, i.e., individual fairness and group fairness \cite{kddSpeicherHGGSWZ18,icseZhangH21,sigsoftHortZSH21}. Individual fairness requires an ML model to produce similar predictive outcomes for similar individuals, while group fairness requires an ML model to treat different groups equally. Since existing bias mitigation methods \cite{nipsWickpT19,icseZhangH21,sigsoftHortZSH21,fairwaypaper,fairsmotepaper,sigsoftBiswasR21} focus on group fairness, in this paper, we also focus on group fairness.

In the context of group fairness, a population is partitioned into the privileged and unprivileged groups based on the values of protected attributes, which refer to the sensitive characteristics (e.g., sex and race) that need to be protected against unfairness. Usually, an unfair ML model tends to favor the privileged group (i.e., inclined to produce the favorable class for its members), thereby putting the unprivileged group at disadvantage. For example, in the recidivism assessment task, race is a protected attribute yet existing recidivism assessment systems have been demonstrated to recommend favourable decisions for white defendants compared to otherwise equivalent black defendants \cite{faircase}.

\subsection{Motivation}
We take the Adult dataset \cite{adultdata}, which is commonly used for predicting income of individuals, as a motivating example to explain what a group fairness issue looks like. In this dataset, 31\% of men and 11\% of women are labeled with ``high income'', a difference of almost three times. As a result, the prediction model trained on this dataset tends to favor men, i.e., it is more likely to predict men as high income than women \cite{fairsmotepaper}. In other words, the obtained prediction model exhibits unfairness regarding sex (a protected attribute in the Adult dataset).

To tackle the group fairness issue, an intuitive idea is to ignore protected attribute information in the training and decision-making process. However, existing work \cite{fairwaypaper} finds that it results in a similar level of unfairness as before. This is because sometimes non-protected attributes employed in the training process may contain information correlated to protected attributes, thereby still leading to potential discrimination.

Researchers have proposed other bias mitigation approaches, including pre-processing, in-processing, and post-processing methods \cite{csurMehrabiMSLG21,DBcorrabs220707068}. 
Pre-processing methods process the training data to mitigate data bias. They take the original training data of ML models as the input, and output the training data with less bias.
In-processing methods improve group fairness during the training process. They take the training data as the input, and are applied in the training process to output fairer models.
Post-processing methods modify the prediction outcomes of ML models to improve fairness. The input is the prediction outcomes of ML models, and the output is fairer outcomes.

These methods can improve fairness at the cost of ML performance. For example, Chakraborty et al. \cite{fairwaypaper} remove ambiguous data points that may contain bias from training data. This helps improve fairness of the obtained ML models, but the loss of information in these data points can result in the decrease of classification accuracy.

Previous studies \cite{biswas2020machine,sigsoftBiswasR21,fairwaypaper,fairsmotepaper} consider both fairness and ML performance in the evaluation of bias mitigation methods. However, the changes in ML performance and fairness caused by bias mitigation methods are often measured and visualised separately. Therefore, it is difficult to judge whether the improved fairness is simply the consequence of ML performance loss. To tackle this problem, Hort et al. \cite{sigsoftHortZSH21} propose a model behavior mutation method to quantitatively benchmark and evaluate the fairness-performance trade-off of different bias mitigation methods.

Nevertheless, existing evaluations of bias mitigation methods have yet to achieve full coverage and completeness to give a comprehensive picture:

\begin{itemize}[leftmargin=*]
\item Existing evaluations are conducted in terms of limited metrics. In particular, researchers often use one or two ML performance metrics for evaluation.
For example, most of the fairness work \cite{icseZhangH21,sigsoftHortZSH21,icdmCaldersKP09,datamineCaldersV10,mfcpaper,kddFeldmanFMSV15,rewpaper,icdmKamiranCP10,isciKamiranMKZ18,aistatsZafarVGG17} measures ML performance using only accuracy.
Chakraborty et al. \cite{fairwaypaper} use recall on the favorable class and false alarm (i.e., 1 minus recall on the unfavorable class) to compare different methods. Biswas and Rajan \cite{biswas2020machine,sigsoftBiswasR21} measure accuracy as well as F1-score on the favorable class. In contrast, Chakraborty et al. \cite{fairsmotepaper} employ the most ML performance metrics, including accuracy and precision/recall/F1-score on the favorable class, but they still ignore other metrics that measure ML performance on the unfavorable class and those that measure overall performance on favorable and unfavorable classes. 

\item Existing evaluations focus on limited bias mitigation methods. For example, Bias and Rajan~\cite{biswas2020machine} evaluate seven of the methods that we use; Chakraborty et al. \cite{fairwaypaper} evaluate five; Chakraborty et al. \cite{fairsmotepaper} evaluate three. Hort et al. \cite{sigsoftHortZSH21} evaluate the most bias mitigation methods, but they focus on only methods proposed in the ML community.  
\end{itemize}

In practice, researchers and practitioners need to choose the bias mitigation method best suited to different application scenarios (i.e., different requirements and evaluation metrics). Existing evaluations cannot provide comprehensive implications for such choices. For evaluation metrics that are not considered by previous studies, engineers have no idea which bias mitigation method is the most suitable. For metrics that have been considered, existing studies use them to evaluate limited methods, so engineers may still have no idea whether the bias mitigation method suggested by the evaluation results is actually the most suitable. For example, because Hort et al. \cite{sigsoftHortZSH21} evaluate only methods from the ML community, software engineers may have no idea whether the methods recently proposed in the SE community have better results. 

To tackle these problems, we present an empirical study with comprehensive measurements and bias mitigation methods from both ML and SE communities. This large-scale study allows us to get a big picture of the literature as well as future research challenges and opportunities. Based on our results, engineers can select the most suitable bias mitigation method for their scenarios according to different requirements (i.e., different evaluation metrics).

While it is true that many practitioners and researchers may have their suspicions about current mitigation approaches, there is no scientific measurement of the effect size of these problems. In this paper, we provide an empirical study to properly measure and understand the magnitude of the problem. This is important also to provide a baseline against which to measure any future improvement. Without this, all we left with is a ``belief'' rather than quantitative scientific evidence.

\section{Experimental Setup}\label{method}
This section describes our research questions and experimental design.

\begin{figure}[!t]
\begin{center}
\includegraphics[width=0.8\columnwidth]{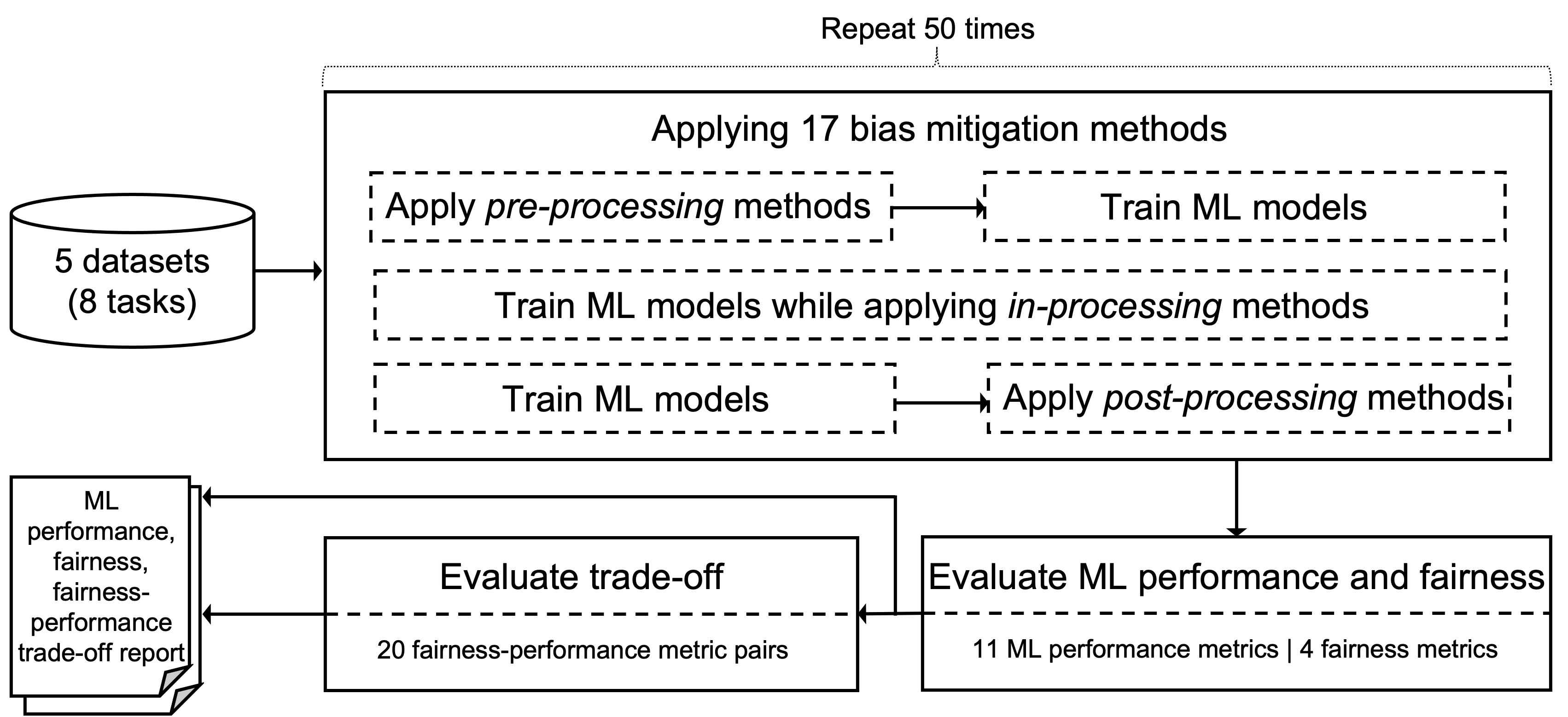}
\caption{Overview of experimental design.}\label{fig:method}
\end{center}
\end{figure}

\subsection{Overview of Experimental Design}
Fig. \ref{fig:method} illustrates our experimental design in a nutshell. First, we use five widely-adopted benchmark datasets (covering eight different software decision tasks) to train ML models. Second, we apply 17 representative bias mitigation methods from the ML and SE communities to these models. Third, we adopt 11 ML performance metrics and 4 fairness metrics to evaluate different methods. Finally, we consider ML performance and fairness together, and evaluate the fairness-performance trade-off of the 17 methods using 20 fairness-performance metric pairs.
Specifically,  we aim to answer the following research questions (RQs):

\textbf{RQ1 (Influence on ML performance):} \emph{How do existing bias mitigation methods affect ML performance?} ML performance (e.g., accuracy and precision) represents important functional requirements of ML software, but bias mitigation methods can improve fairness at the cost of ML performance. Therefore, we first use various metrics to investigate how existing methods change ML performance.

\textbf{RQ2 (Influence on fairness):} \emph{How do existing bias mitigation methods affect fairness?} Since the main goal of bias mitigation methods is to improve software fairness, we then use various fairness metrics to measure how well existing bias mitigation methods achieve this goal.

\textbf{RQ3 (Fairness-performance trade-off):} \emph{What fairness-performance trade-off do existing bias mitigation methods achieve?} We finally consider fairness and ML performance together, and evaluate existing bias mitigation methods using different types of fairness-performance trade-off assessment, i.e., combinations of different fairness metrics and ML performance metrics.

In the following, we introduce the bias mitigation methods (Section \ref{17methods}), benchmark datasets (Section \ref{benchdata}), metrics and measures (Section \ref{metricmeasure}), and experimental settings (Section \ref{settings}).

\subsection{Bias Mitigation Methods}\label{17methods}
We focus our analysis on 17 representative bias mitigation methods proposed in the ML and SE communities. 
For the methods in the ML community, we follow previous work \cite{biswas2020machine,icseZhangH21,sigsoftHortZSH21,fairwaypaper} to employ the state-of-the-art ones implemented in the IBM AIF360 framework \cite{aif360}. Specifically, we employ all the ten methods listed on its homepage~\cite{aif360}, covering three types, i.e., pre-processing, in-processing, and post-processing methods. 
Next, we briefly introduce each method by type.

\emph{Pre-processing:} Optimized Pre-processing (\textbf{OP})~\cite{oppaper} learns a probabilistic transformation to modify data features and labels. Learning Fair Representation (\textbf{LFR})~\cite{LFRpaper} learns fair representations by obfuscating information about protected attributes. Reweighting (\textbf{RW})~\cite{rewpaper} generates different weights for samples in each (group, label) combination. Disparate Impact Remover (\textbf{DIR}) \cite{DIpaper} modifies feature values to improve fairness while preserving rank-ordering within groups.

\emph{In-processing:} Prejudice Remover (\textbf{PR})~\cite{PRpaper} adds a discrimination-aware regularization term to the learning objective. Adversarial Debiasing (\textbf{AD})~\cite{ADVpaper} uses adversarial techniques to maximize accuracy and reduce evidence of protected attributes in the predictions simultaneously. Meta Fair Classifier (\textbf{MFC})~\cite{mfcpaper} takes the fairness metric as part of the input and returns a classifier optimized for the metric.

\emph{Post-processing:} Reject Option Classification (\textbf{ROC})~\cite{ROCpaper} targets predictions with high uncertainty and tends to assign favorable outcomes to the unprivileged group and unfavorable outcomes to the privileged group. Calibrated Equalized Odds Post-processing (\textbf{CEO})~\cite{COpaper} optimizes over calibrated classifier score outputs to find probabilities with which to change output labels with an equalized odds objective. Equalized Odds Post-processing (\textbf{EOP})~\cite{EOpaper} solves a linear program to find probabilities with which to change output labels to optimize equalized odds.

As for the methods proposed in the SE community, we use two methods recently published in top SE venues, including Fairway \cite{fairwaypaper} at ESEC/FSE 2020 and Fair-SMOTE \cite{fairsmotepaper} at ESEC/FSE 2021. 

\textbf{Fairway}~\cite{fairwaypaper} combines pre-processing and in-processing techniques to improve fairness. First, it evaluates the original labels of the training data and removes ambiguous data points that can eventually make the classifier biased. Then, it employs multi-objective optimization to maximize the model performance while making it fair. 

\textbf{Fair-SMOTE}~\cite{fairsmotepaper} is a pre-processing method that employs two strategies. First, it generates new data points to make the numbers of training data in different subgroups (i.e., combinations of different outcomes and protected attribute values) equal. Second, it uses the same method as Fairway to remove ambiguous data points from the training data.

In the AIF360 toolkit, MFC, ROC, and CEO are implemented with two, three, and three different metrics to guide the bias mitigation process, respectively. Specifically, MFC offers a choice between Disparate Impact (DI) and False Discovery Rate (FDR); ROC offers a choice among Statistical Parity Difference (SPD), Average Odds Difference (AOD), and Equal Opportunity Difference (EOD); CEO offers a choice among False Negative Rate (FNR), False Positive Rate (FPR), and a weighted metric to combine both. We implement and evaluate each of the settings. Therefore, we have a total of 17 bias mitigation methods for our study.

\subsection{Benchmark Datasets}\label{benchdata}
We follow previous work \cite{icseZhangH21} to use five benchmark datasets implemented in the IBM AIF360 (as listed in Table~\ref{dataset_info}). The five datasets cover social, financial, and medical domains, and are widely adopted in the fairness literature~\cite{csurMehrabiMSLG21,icseZhangH21,sigsoftHortZSH21,fairwaypaper,fairsmotepaper,sigsoftBiswasR21}. The number of datasets used in this study compares favorably with the literature, as previous work \cite{sigsoftHortZSH21} points out that 90\% of fairness papers use no more than three datasets. Next, we briefly introduce each dataset.

\begin{table}[!tp]
\centering
\small
\caption{Benchmark datasets for bias mitigation.}
\label{dataset_info}
\begin{tabular}{lrllr}
\hline
Name & Size  & Protected attribute(s) & Favorable label & Majority label\\
\hline
Adult & 45,222 & Sex, Race & 1 (income $\textgreater$ 50K)  & 0 (75.2\%)\\
Compas & 6,167 & Sex, Race & 0 (no recidivism) & 0 (54.5\%) \\
German & 1,000  & Sex, Age & 1 (good credit) & 1 (70.0\%) \\
Bank & 30,488  & Age  & 1 (subscriber)& 0 (87.3\%)\\
Mep & 15,830 &Race  & 1 (utilizer)& 0 (82.8\%)\\
\hline
\end{tabular}
\end{table}

Adult Income dataset \cite{adultdata} (a.k.a., \textbf{Adult} dataset) contains demographic and financial information about individuals extracted from the 1994 U.S. census data, and is used to predict whether the income of a person exceeds \$50K a year or not.

ProPublica Recidivism dataset \cite{compasdata} (a.k.a., \textbf{Compas} dataset) contains demographic information and criminal histories of defendants from Broward County, and is used to predict whether a defendant will be re-offended within two years. 

German Credit dataset \cite{germandata} (a.k.a., \textbf{German} dataset) contains demographic and credit information of 1,000 individuals, and is used to predict people's credit risk levels.

Bank Marketing dataset \cite{bankdata} (a.k.a., \textbf{Bank} dataset) contains demographic, social, and financial information of clients of a Portuguese banking institution, and is used to predict whether a client will subscribe a term deposit.

Medical Survey 2015 dataset \cite{mepdata} (a.k.a., \textbf{Mep} dataset) contains data measuring how Americans use and pay for medical care, health insurance, and out-of-pocket spending, and is used to predict health care utilization of individuals.

As shown in Table~\ref{dataset_info}, each dataset has its protected attribute(s) specified by its provider. We acknowledge that there may be other attributes that also need to be protected for each dataset. However, determining all protected attributes needs the assistance of requirements engineers, even legal practitioners, compliance officers, and policy makers, which is beyond the scope of this paper. Therefore, we follow previous work \cite{icseZhangH21,sigsoftHortZSH21,fairwaypaper,sigsoftBiswasR21} to use the specified protected attributes for study. In addition, although in practice software engineers may need to mitigate bias regarding all potentially protected attributes, most of existing bias mitigation methods treat each protected attribute individually for each task. Therefore, in line with previous work~\cite{icseZhangH21,sigsoftHortZSH21,fairwaypaper,sigsoftBiswasR21}, we consider one protected attribute each time and thus have eight dataset-attribute pairs (e.g., Adult-Sex and Adult-Race). We use the eight pairs as the eight bias mitigation tasks of this study and take each task as a binary classification problem.

\subsection{Evaluation Metrics and Measures}\label{metricmeasure}
We use 11 ML performance metrics, 4 fairness metrics, and 20 types of fairness-performance trade-off assessment for a comprehensive evaluation of existing bias mitigation methods. 
In the following, we briefly introduce the metrics and measures that we use.

Given a bias mitigation task, let a protected attribute be $A$, with 0 as the unprivileged group and 1 the privileged group; let the real classification label be $Y$ and the predicted label $\hat{Y}$, with 0 as the unfavorable class and 1 the favorable class. In addition, we use $Pr$ to denote probability.

\subsubsection{ML Performance Metrics}\label{performance_metric}
For each bias mitigation method, we measure the ML performance changes caused by it on favorable and unfavorable classes in terms of precision, recall, and F1-score, which are widely employed in classification problems \cite{NovielliGL08,tosemChenCYLPML21,sigsoftChenCLML19,msrNovielliCDGL20}.

\textbf{Precision} measures the exactness of a method. The precision for a given class \textit{c} (i.e., 0 or 1) is calculated as:
\begin{equation}
Precision@c = Pr[Y=c|\hat{Y}=c].
\end{equation}

\textbf{Recall} measures the sensitivity of a method. The recall for a given class \textit{c} is calculated as:
\begin{equation}
Recall@c = Pr[\hat{Y}=c|Y=c].
\end{equation}

\textbf{F1-score} measures a harmonic mean of precision and recall. The F1-score for a given class \textit{c} is calculated as:

\begin{equation}
F1@c = \frac{2 \times Precision@c \times Recall@c}{Precision@c+Recall@c}.
\end{equation}

%https://arxiv.org/abs/1505.01658
In addition, to measure the overall performance, we follow previous classification work in SE \cite{sigsoftHortZSH21,icseZhangH21,fairsmotepaper,biswas2020machine,sigsoftBiswasR21} to use accuracy (\textbf{Acc}), which measures how often a method makes the correct prediction and is calculated as:

\begin{equation}
Acc = Pr[\hat{Y} = Y].
\end{equation}

Acc is often criticized as not being suitable for the imbalanced class distribution, because it is easy for an ML model to obtain a high accuracy just by predicting all samples as the majority class in such a distribution.
Considering that some datasets (e.g., the Bank dataset) that we use have an imbalanced class distribution, we follow previous SE work to use three additional macro-average metrics  \cite{NovielliGL08,msrNovielliCDGL20,tosemChenCYLPML21} and the Matthews Correlation Coefficient (\textbf{MCC}) metric \cite{easeYaoS20,RodriguezHHDR14}, which are all demonstrated to be suitable for dealing with imbalanced scenarios~\cite{SebastianiMachine,chicco2020advantages}. For the macro-average metrics, we use macro-precision (\textbf{Mac-P}), macro-recall (\textbf{Mac-R}), and macro-F1 (\textbf{Mac-F1}), which take the average of precision, recall, and F1-score on the favorable and unfavorable classes, respectively. MCC is calculated as:

\begin{equation}
MCC = \frac{TP \times TN-FP \times FN}{\sqrt{(TP+FP)(TP+FN)(TN+FP)(TN+FN)}},
\end{equation}

where TP, TN, FP, and FN denote the numbers of favorable samples predicted as favorable, unfavorable samples predicted as unfavorable, unfavorable samples predicted as favorable, and favorable samples predicted as unfavorable, respectively.

To summarize, we use 11 ML performance metrics, including F-P (precision for the favorable), F-R (recall for the favorable), F-F1 (F1-score for the favorable), UnF-P (precision for the unfavorable), UnF-R (recall for the unfavorable), UnF-F1 (F1-score for the unfavorable), Acc, Mac-P, Mac-R, Mac-F1, and MCC. The values of F-P, F-R, F-F1, UnF-P, UnF-R, UnF-F1, Acc, Mac-P, Mac-R, and Mac-F1 are between 0 and 1. The value of MCC is between -1 and 1, where 1 represents a perfect prediction, 0 no better than random prediction, and -1 total disagreement between prediction and observation. For all the metrics, larger values indicate better ML performance.

% Although accuracy, precision, recall, and F1-score are generally more representative metrics, 
In practice, the choice of metrics depends on the intended applications. Specifically, requirements engineers can determine the metrics suitable for their applications without the need of considering all the 11 metrics. For example, for disease detection systems, engineers may pursue a high recall on the unfavorable class (i.e., disease). 
Indeed, different types of datasets have different appropriate metrics. In our study, we use the full set of metrics for all the datasets due to the following reasons: 1) in practice, it is difficult to predict which metric is required, and thus using the full set of metrics in the study provides more comprehensive information for engineers to get reference when choosing metrics; 2) existing bias mitigation methods have used different types of metrics based on inconsistent rules, and thus using the full set of metrics for different datasets and methods provides a comprehensive and unified way to make comparisons.

\subsubsection{Fairness Metrics}\label{fairnessmetrics}
Based on different definitions of fairness, various fairness metrics have been proposed to measure the difference in classification between the privileged and unprivileged groups. 
In this work, we use the group fairness metrics that are most widely adopted in fairness research \cite{biswas2020machine,sigsoftBiswasR21,icseZhangH21,sigsoftHortZSH21,fairwaypaper,fairsmotepaper}.

Statistical Parity Difference (\textbf{SPD}) measures the difference in the acceptance rate of the favorable class between the privileged and unprivileged groups:

\begin{equation}
\begin{aligned}
SPD = Pr[\hat{Y} = 1 | A = 0] - Pr[\hat{Y} = 1 | A = 1].
\end{aligned}
\end{equation}

Average Odds Difference (\textbf{AOD}) measures the average of differences in the false positive rate and the true positive rate between the privileged and unprivileged groups:

\begin{equation}
\begin{aligned}
AOD  = &\frac{1}{2}(\vert Pr[\hat{Y}=1| A=0, Y=0]  -  Pr[\hat{Y}=1|A=1, Y=0]\vert \\ & + \vert Pr[\hat{Y}=1| A=0, Y=1]  -  Pr[\hat{Y}=1|A=1, Y=1] \vert).
\end{aligned}
\end{equation}

Equal Opportunity Difference (\textbf{EOD}) measures the difference in the true positive rate between the privileged and unprivileged groups:

\begin{equation}
\begin{aligned}
EOD & = Pr[\hat{Y}=1| A=0, Y=1]  -  Pr[\hat{Y}=1|A=1, Y=1].
\end{aligned}
\end{equation}

Error Rate Difference (\textbf{ERD}) measures the difference in the error rate between  the privileged and unprivileged groups:

\begin{equation}
\begin{aligned}
ERD & = Pr[\hat{Y} \neq Y|A=0] - Pr[\hat{Y} \neq Y|A=1].
\end{aligned}
\end{equation}

There is another fairness metric called Disparate Impact (DI), which is also widely adopted in the fairness literature. DI and SPD both compare the probabilities of classifying samples as favorable in the privileged and unprivileged groups. Specifically, DI computes the ratio of the two probabilities, while SPD computes the difference of the two probabilities.
% When computing DI, we may encounter the divided-by-zero error. 
Between SPD and DI, we follow previous work \cite{sigsoftBiswasR21,sigsoftHortZSH21} to use only SPD in our evaluation. In practice, the different acceptance rates of the favorable class between the privileged and unprivileged groups are allowed in certain scenarios, where SPD may be not a good metric. When it comes to fairness research, SPD is often adopted despite the existence of such threats because the main purpose of fairness research is to investigate the effectiveness of bias mitigation methods, in which scenario the validity of the bias in the data does not affect the evaluation of bias mitigation methods.

For all the fairness metrics, we use their absolute values. In this way, values equal to 0 indicate the greatest fairness; larger values indicate more bias. 

\subsubsection{Fairness-performance Trade-off Measures}\label{faireades}
It is difficult to evaluate which bias mitigation method is better based on fairness alone, since it is unclear whether the improved fairness is simply the consequence of ML performance loss. Therefore, here, we consider fairness and ML performance together and measure the fairness-performance trade-off of different methods. 

To this end, we employ Fairea \cite{sigsoftHortZSH21}, a model behavior mutation method proposed at ESEC/FSE 2021, to benchmark and quantify the fairness-performance trade-off achieved by existing bias mitigation methods. 
Next, we briefly introduce Fairea, which is illustrated in Fig. \ref{fig:fairea}.\footnote{The figure is taken from the original paper \cite{sigsoftHortZSH21}.} Specifically, Fairea includes three steps as follows:

\begin{figure} 
    \centering
       \includegraphics[width=0.8\linewidth]{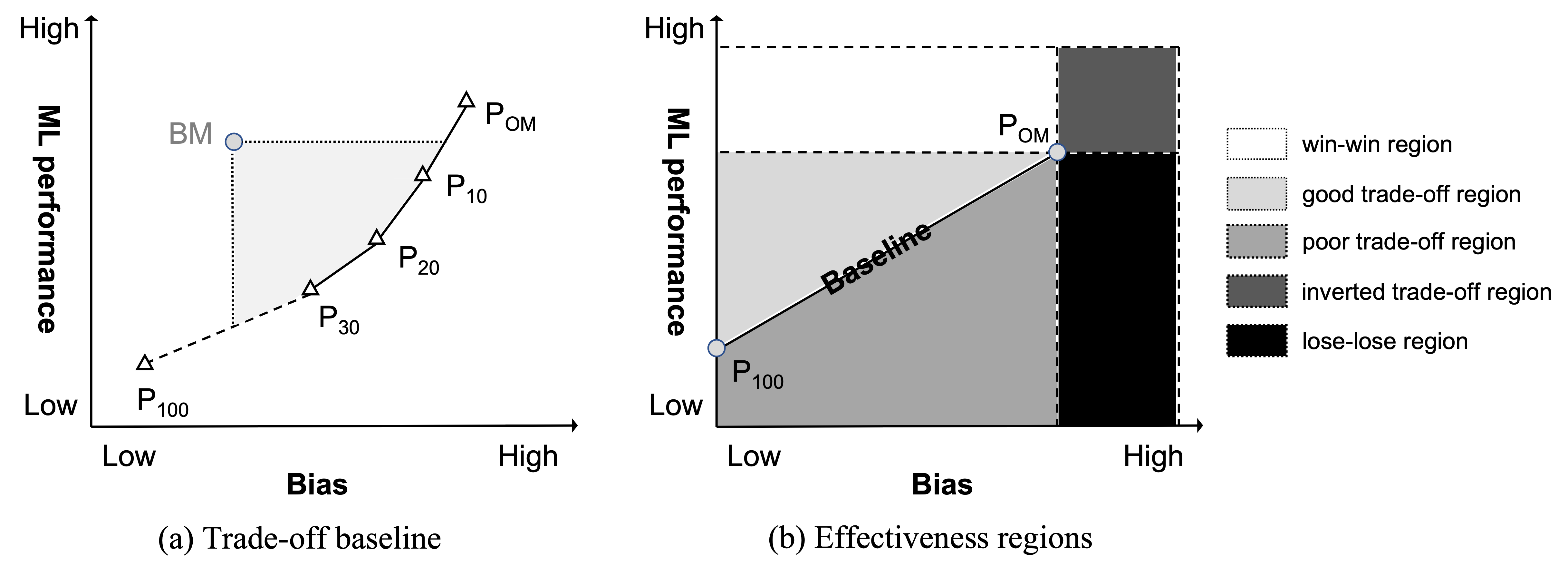}
  \caption{Illustration of Fairea \cite{sigsoftHortZSH21}. (a) presents the fairness-performance trade-off baseline, where $P_{OM}$ represents the original model, BM the model after bias mitigation, and  $P_{10}, ..., P{100}$ the points obtained by model behavior mutation. (b) presents the effectiveness regions of bias mitigation methods according to the changes in ML performance and bias.}
  \label{fig:fairea} 
\end{figure}

\emph{Step 1: Create trade-off baseline.} Fairea presents the ML performance and fairness achieved by bias mitigation methods in a two-dimensional coordinate system as shown in Fig. \ref{fig:fairea}(a). It constructs the fairness-performance trade-off baseline by connecting the original model (i.e., $P_{OM}$) and a series of pseudo mitigation models generated by model behavior mutation (i.e., $P_{10}$, ..., $P_{100}$). The pseudo mitigation models are mutated based on the original model by sacrificing ML performance to reduce bias in a naive way. Specifically, Fairea randomly chooses a subset of the predictions made by the original model and replaces them with the majority class of the data. It considers different mutation degrees (i.e., the fraction of chosen predictions) from 10\% to 100\%, with a step size of 10\%, to obtain $P_{10}$, ..., $P_{100}$. The core insight of Fairea is that when the original model is mutated into a model that always predicts the same class, fairness will be greatly improved as predictive performance is equally worse in the privileged and unprivileged groups. The fairness-performance trade-off of these naive mutated models is expected to be surpassed by any reasonable bias mitigation method, so we use these models as the baseline. 

\emph{Step 2: Divide effectiveness regions.} The obtained baseline categorizes bias mitigation methods into five regions that represent different effectiveness levels. As illustrated in Fig. \ref{fig:fairea}(b), the \textbf{win-win region} contains bias mitigation methods that improve ML performance and decrease bias, while the \textbf{lose-lose region} contains methods that decrease ML performance and increase bias. Methods that improve both ML performance and bias fall in the \textbf{inverted trade-off region}. The remaining two regions contain methods that reduce bias but decrease ML performance. Specifically, if a method achieves a better trade-off than the baseline constructed by Fairea, it falls within the \textbf{good trade-off region}. Otherwise, it belongs to the \textbf{poor trade-off region}. The region division of Fairea provides an overview of the overall effectiveness of a bias mitigation method.

\emph{Step 3: Quantify trade-off effectiveness.} Fairea can also measure the trade-off effectiveness of a bias mitigation method in a quantitative way. Fig. \ref{fig:fairea}(a) shows the area (indicated in grey) obtained by connecting the point of the model after applying a bias mitigation method (i.e., \textit{BM}) and the Fairea baseline vertically and horizontally. Fairea calculates the size of the area as a quantitative measure of the fairness-performance trade-off achieved by the bias mitigation method. A larger area indicates a better trade-off. Using the area as a measure of the trade-off enables a convenient comparison among different bias mitigation methods.

In the original paper \cite{sigsoftHortZSH21}, Fairea evaluates only two types of fairness-performance trade-offs (i.e., SPD\&Acc and AOD\&Acc) on 12 bias mitigation methods proposed in the ML community. In this study, we aim to extend our evaluation to the trade-off between more fairness and performance metrics on 17 bias mitigation methods from the ML and SE communities. Since we employ 11 ML performance metrics and 4 fairness metrics, we have a total of 48 fairness-performance metric pairs. However, as Fairea replaces predictions with the majority class, in terms of ML performance metrics for a certain class (e.g., recall on the majority class), we may observe that ML performance and fairness are both improved with the increased mutation degrees. For such metrics, we cannot obtain the trade-off baselines as in Fig. \ref{fig:fairea}(a). Therefore, here we choose only 5 ML performance metrics that measure performance on both the favorable and unfavorable classes, i.e., Acc, Mac-P, Mac-R, Mac-F1, and MCC, to reflect the overall performance of each bias mitigation method. As a result, we have 20 types of fairness-performance trade-offs, i.e., combinations of 5 ML performance metrics and 4 fairness metrics. 

\subsection{Experimental Settings}\label{settings}
To ensure the reproducibility of our study, we describe the experimental settings in details.

\textbf{Implementation of datasets:} 
We use the five benchmark datasets implemented in the IBM AIF360 via directly invoking off-the-shelf APIs provided by it \cite{aif360}. Moreover, we follow previous work \cite{fairsmotepaper,fairwaypaper,icseZhangH21,sigsoftHortZSH21} to normalize all feature values to be between 0 and 1.

\textbf{Implementation of bias mitigation:} For each bias mitigation task, we train original models using three traditional ML algorithms and four Deep Neural Networks (DNNs): 
\begin{itemize}[leftmargin=*]
\item For traditional ML algorithms: We use Logistic Regression (LR), Support Vector Machine (SVM), and Random Forest (RF), all of which are widely adopted in previous studies \cite{sigsoftHortZSH21,icseZhangH21,biswas2020machine}. In line with these studies, we use the default configuration provided by the scikit-learn library~\cite{sklearn} to implement each algorithm.
\item For DNNs: We first use a fully-connected neural network composed of five hidden layers, which contain 64, 32, 16, 8, 4 units, respectively. This DNN is widely applied to our benchmark datasets in previous fairness research \cite{mengdi22,icseZhangW0D0WDD20,icseZhengCD0CJW0C22}. For a more comprehensive evaluation, we also use another three fully-connected neural networks used in a previous fairness testing study \cite{isstaZhangZZ21}. The hidden layer structures of the three DNNs are [50, 30, 15, 10, 5], [30, 20, 15, 10, 5], and [30, 20, 15, 15, 10], respectively. In line with these previous studies, we use ReLU and Softmax as the activation functions for the hidden layers and the output layer, respectively. In the rest of this paper, we denote the four DNNs as DL1, DL2, DL3, and DL4, respectively.
\end{itemize}
 
We apply 17 bias mitigation methods for the original models. Specifically, pre-processing and post-processing methods are applied before and after model training, while in-processing methods are applied during the training process to obtain new models.
We implement the 15 bias mitigation methods from the ML community based on the IBM AIF360 framework \cite{aif360}, and implement the two methods from the SE community based on the code released by their authors \cite{fairwaygit,fairsmotegit}. 
Since the IBM AIF360 does not support the OP method for the Bank and Mep datasets, we apply OP only for six tasks. For the remaining 16 methods, we apply each of them for all the eight bias mitigation tasks. Each application is repeated 50 times. Each time, the dataset is shuffled and randomly splitted into 70\% training data and 30\% test data. 

\textbf{Implementation of Fairea:} We use Fairea to create the fairness-performance trade-off baseline for each \emph{(task, model, fairness-performance metric pair)} combination. Specifically, we train the original model 50 times. Each time, based on the original model, we repeat the mutation procedure 50 times for each mutation degree, i.e., 10\%, 20\%, ..., and 100\%. Finally, as suggested by Fairea \cite{sigsoftHortZSH21}, we construct the baseline using the mean value of the multiple runs.

\textbf{Statistical analysis:} To test whether the difference between two bias mitigation methods is statistically significant, we employ the non-parametric Mann Whitney U-test \cite{mann1947test}. This test suits our purpose well as it does not assume normality. The difference is considered significant, only if the $p$-value of the computed statistic is lower than a pre-specified level (usually 0.05). For example, when we use the Mann Whitney U-test to compare two sets of accuracy values achieved by the 50 runs of methods \emph{A} and \emph{B}, the null hypothesis is that the accuracy of \emph{A} is similar to \emph{B}. If we find that $p$-value $\textless$ 0.05, we can conclude with 95\% confidence that our alternative hypothesis is true, which indicates that \emph{A} achieves a significantly different accuracy than \emph{B}. By default, we use 95\% as the confidence level in this paper.

Furthermore, we compute the effect size with the Cohen’s $d$ \cite{cohen2013statistical}, to check whether the difference has a meaningful effect.
We consider the difference with 0 $\textless$ $d$ $\textless$ 0.5 a small effect, 0.5 $\leq$ $d$ $\textless$ 0.8 a medium effect, and $d$ $\geq$ 0.8 a large effect \cite{sawilowsky2009new}. 

In addition, we employ the Spearman's rank correlation coefficient $\rho$ \cite{myers2013research} to investigate whether changes in the values of different metrics caused by bias mitigation methods are similar. Spearman's rank correlation coefficient does not assume normality, and thus suits our purpose. The value of $\rho$ is between -1 to 1, where -1 indicates perfectly negative correlation, 0 no correlation, 1 perfectly positive correlation. Moreover, the $p$-value is reported together with the correlation coefficient. The correlation is considered statistically significant, only if the $p$-value is lower than 0.05.

\textbf{Experimental environment:} The experiments are implemented with Python 3.7.11 and TensorFlow 1.15.0, and executed on a Ubuntu 16.04 LTS with 128GB RAM, having 2.3 GHz Intel Xeon E5-2653 v3 Dual CPU and two NVidia Tesla M40 GPUs.

\section{Results}\label{results}
In this section, we answer our RQs based on experimental results.

\begin{figure}[!t]
\begin{center}
\includegraphics[width=0.95\columnwidth]{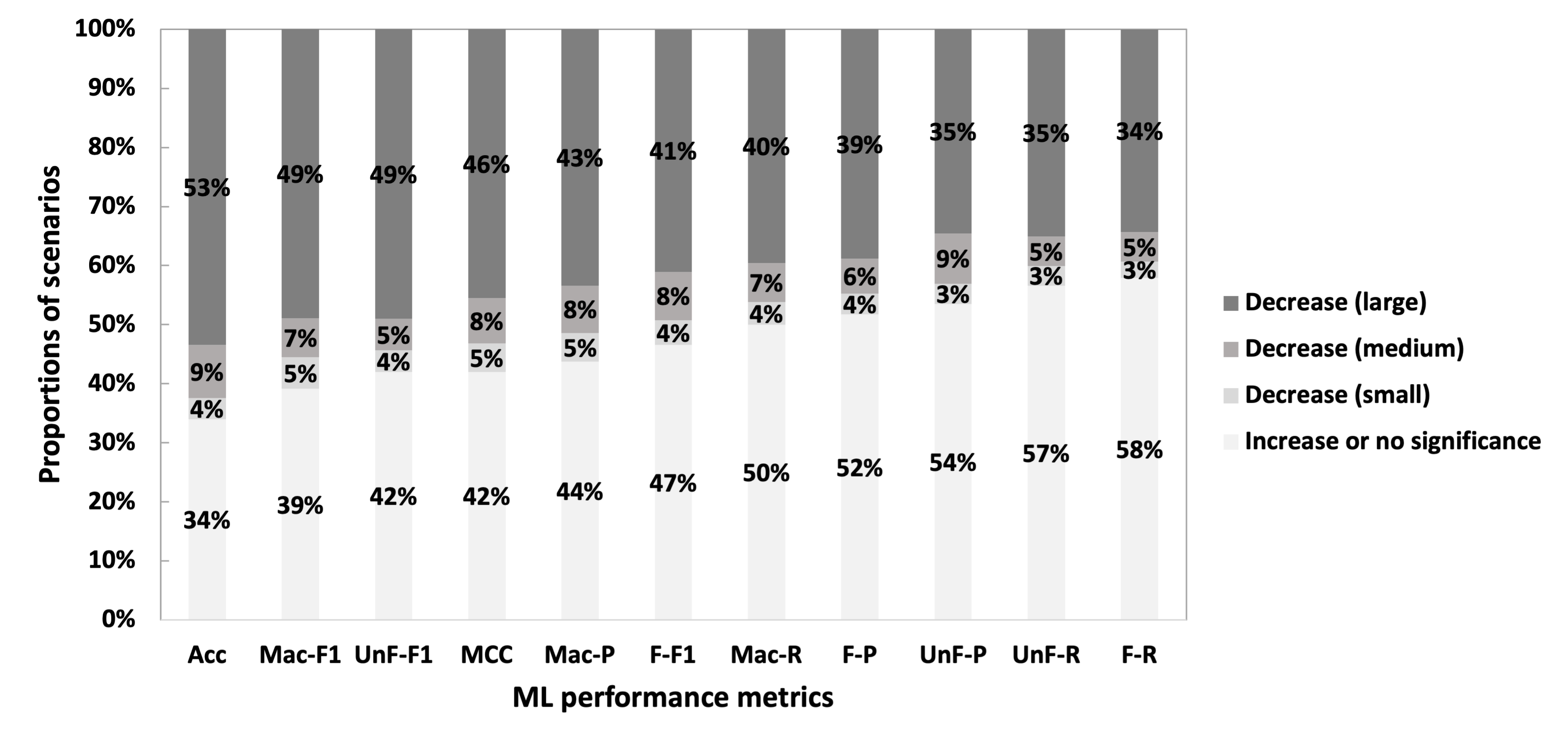}
\caption{\textbf{RQ1.1:} Effects of bias mitigation methods on different ML performance metrics. After applying existing bias mitigation methods, the values of the 11 ML performance metrics significantly decrease in an average of 53\% of applications.}\label{fig:effect_per}
\end{center}
\end{figure}

\subsection{RQ1: Influence on ML Performance}\label{perform_result}
This section presents the results for 11 ML performance metrics. Based on the results, we investigate RQ1 by answering two specific questions:

\textbf{RQ1.1 (Effects on ML performance metric values):} \emph{How do the values of ML performance metrics change after bias mitigation?} First, we investigate whether the values of ML performance metrics are significantly changed after we apply existing bias mitigation methods. Additionally, we explore whether the changes in different ML performance metrics are significantly correlated. If so, researchers and practitioners can employ the ML performance metrics used in previous work as proxies for the unconsidered ones.

\textbf{RQ1.2 (ML performance comparison among methods):} \emph{How do different bias mitigation methods affect ML performance?} Second, we compare existing bias mitigation methods in terms of different ML performance metrics. The results provide implications for the choice of bias mitigation methods in application scenarios where ML performance is critical.

\subsubsection{RQ1.1: Effects on ML Performance Metric Values}\label{performance_effect}
To answer RQ1.1, we use the original models that do not apply any bias mitigation method as the baselines. Specifically, for each bias mitigation task, we use LR, SVM, RF, DL1, DL2, DL3, and DL4 to train the original models, with each model repeated 50 times. Then, for each task-model pair, we use the average level (i.e., mean values of ML performance metrics) of the corresponding 50 original models as the baseline. Since we apply each bias mitigation method in each task-model pair, we have a total of $16\times8\times7+1\times6\times7 = 938$ applications.\footnote{Note that we apply the OP method in six tasks and the other 16 methods in eight tasks, as described in Section \ref{settings}.} We repeat each application 50 times. For each application, we calculate the difference in the mean value of each ML performance metric achieved by the 50 models after applying the bias mitigation method and the corresponding 50 original models. Then we analyze the significance and effect size of the difference using Mann Whitney U-test and Cohen's $d$. In Fig. \ref{fig:effect_per}, for each ML performance metric, we present the proportions of applications that fall into different effects, i.e., decreasing ML performance significantly ($p$-value $\textless$ 0.05) with a large effect ($d$ $\geq$ 0.8), decreasing ML performance significantly with a medium effect (0.5 $\leq$ $d$ $\textless$ 0.8), decreasing ML performance significantly with a small effect (0 $\textless$ $d$ $\textless$ 0.5) , and increasing ML performance or decreasing ML performance insignificantly ($p$-value $\geq$ 0.05). 

From Fig. \ref{fig:effect_per}, we observe that the values of the 11 ML performance metrics decrease significantly in an average of 53\% of applications (ranging from 42\% to 66\% according to different metrics). This observation is in line with the intuition that lessening the impacts of protected attributes is likely to lessen ML performance. In particular, the values of the 11 metrics significantly decrease with a large effect in an average of 42\% of applications (ranging from 34\% to 53\%). However, as described in Section \ref{intro}, existing studies often evaluate bias mitigation methods in terms of only one or two ML performance metrics, ignoring some important metrics such as UnF-P and UnF-F1. As a result, researchers and practitioners may be unaware of the significantly decreased performance caused by bias mitigation methods in these unconsidered metrics and thus choose inappropriate methods, making the functional properties of ML software not up to expectations. 
% For example, the PR method achieves high F-P, but low UnF-P. However, existing evaluations do not take UnF-P into consideration. If practitioners use the PR method in online advertising tasks that target low-income users and care about UnF-P (as discussed in Section \ref{intro}), they may obtain unsatisfactory results.

\begin{table}[!tp]
\setlength\tabcolsep{1.1pt}
\centering
\renewcommand\arraystretch{1}
\tiny
\caption{\textbf{RQ1.1:} Correlation between ML performance metrics. In the table, metrics that have been considered by previous work are highlighted in shading; * indicates a significant correlation overall ($p$-value $\textless$ 0.05); numbers in parentheses indicate that in how many task-model pairs the correlation shares the consistent pattern with the overall correlation. We observe that the effects of bias mitigation methods on the ML performance metrics newly considered in this paper do not have a consistent correlation with previously employed metrics across all the task-model pairs.}
\label{corrresult}
\begin{tabular}{l|rrrrrrrrrrr}
\hline
  & \cellcolor{gray!25}F-R & \cellcolor{gray!25}F-F1 & UnF-P & \cellcolor{gray!25}UnF-R & UnF-F1 & \cellcolor{gray!25}Acc & Mac-P & Mac-R & Mac-F1 & MCC\\
\hline
\cellcolor{gray!25}F-P& -0.573*(38/56)& -0.139*(6/56)& -0.328*(15/56)& 0.719*(54/56)& 0.777*(54/56)& 0.589*(30/56)& 0.732*(39/56)& 0.056(0/56)& 0.418*(35/56)& 0.318*(26/56)\\
\cellcolor{gray!25}F-R& -& 0.727*(51/56)& 0.752*(42/56)& -0.888*(56/56)& -0.588*(36/56)& -0.052(27/56)& -0.154*(5/56)& 0.446*(30/56)& 0.169*(12/56)& 0.281*(18/56)\\
\cellcolor{gray!25}F-F1& -& -& 0.608*(47/56)& -0.520*(40/56)& -0.158*(16/56)&  0.412*(27/56)& 0.264*(22/56)& 0.570*(32/56)& 0.598*(32/56)& 0.672*(34/56)\\
UnF-P& -& -& -& -0.571*(31/56)& -0.242*(9/56)& 0.247*(28/56)& 0.221*(28/56)& 0.615*(38/56)& 0.398*(25/56)& 0.519*(36/56)\\
\cellcolor{gray!25}UnF-R& -& -& -& -& 0.831*(53/56)& 0.228*(15/56)& 0.306*(21/56)& -0.102*(27/56)& 0.162*(14/56)& 0.056(10/56)\\
UnF-F1& -& -& -& -& -& 0.531*(29/56)& 0.523*(33/56)& 0.266*(18/56)& 0.553*(32/56)& 0.453*(19/56)\\
\cellcolor{gray!25}Acc&  -& -& -& -& -& -& 0.881*(56/56)& 0.267*(19/56)& 0.666*(37/56)& 0.604*(37/56)\\
Mac-P& -&  -& -& -& -& -& -& 0.194*(19/56)& 0.558*(31/56)& 0.522*(31/56)\\
Mac-R& -& -& -& -& -& -& -& -& 0.798*(49/56)& 0.853*(55/56)\\
Mac-F1& -&  -& -& -& -& -& -& -& -& 0.959*(56/56)\\
\hline
\end{tabular}
\end{table}

Next, we calculate the Spearman's rank correlation coefficient $\rho$ between every pair of ML performance metrics. Specifically, we calculate the value differences of each ML performance metric before and after bias mitigation in the 938 applications. Then, for each two ML performance metrics, we calculate the overall correlation between the 938 value differences of the two metrics. In addition, we also calculate the correlation in the $8\times7 = 56$ task-model pairs, separately. Table \ref{corrresult} shows the correlation results. For each metric pair, we present the overall $\rho$, with * indicating a significant correlation (i.e., $p$-value $\textless$ 0.05). Additionally, we list the number of task-model pairs where the correlation shares the consistent pattern with the overall correlation in parentheses. For example, the correlation result of F-P and MCC is 0.318*(26/56), indicating that the value differences of F-R and MCC have a significantly positive correlation overall, and the significantly positive correlation holds in 25 out of 56 task-model pairs. Overall, among the 55 metric pairs in Table \ref{corrresult}, 41 show a significantly positive correlation, 11 a significantly negative correlation, 3 no significant correlation. Furthermore, we find that 52 metric pairs do not present a consistent correlation pattern across all the task-model pairs. Specifically, only the (Mac-F1, MCC), (Acc, Mac-P), and (F-R, UnF-R)  pairs present a consistent correlation pattern. However, not all the metrics are considered in previous software fairness work \cite{sigsoftHortZSH21,icseZhangH21,fairsmotepaper,fairwaypaper,biswas2020machine,sigsoftBiswasR21}. 
Since the metrics newly considered in this work are not consistently correlated  with any previously employed metric, we may not use the latter as their proxy. This finding suggests that we take comprehensive ML performance metrics into account during the evaluation of bias mitigation methods, especially considering that different ML performance metrics measure the functional properties of ML software from different aspects and thus may provide different guidelines for real-world application scenarios.

\finding{\textbf{Finding 1:} The values of all the 11 ML performance metrics (including those not considered in previous work) decrease significantly in a notable proportion of applications (42\%$\sim$66\% according to different metrics) after applying existing bias mitigation methods. In particular, these methods significantly decrease accuracy in 66\% of scenarios. Moreover, the effects of bias mitigation methods on the ML performance metrics newly considered in this work do not consistently correlated with previously employed metrics, and therefore the latter cannot be used as a proxy.}

\begin{figure}[!t]
\begin{center}
\includegraphics[width=1\columnwidth]{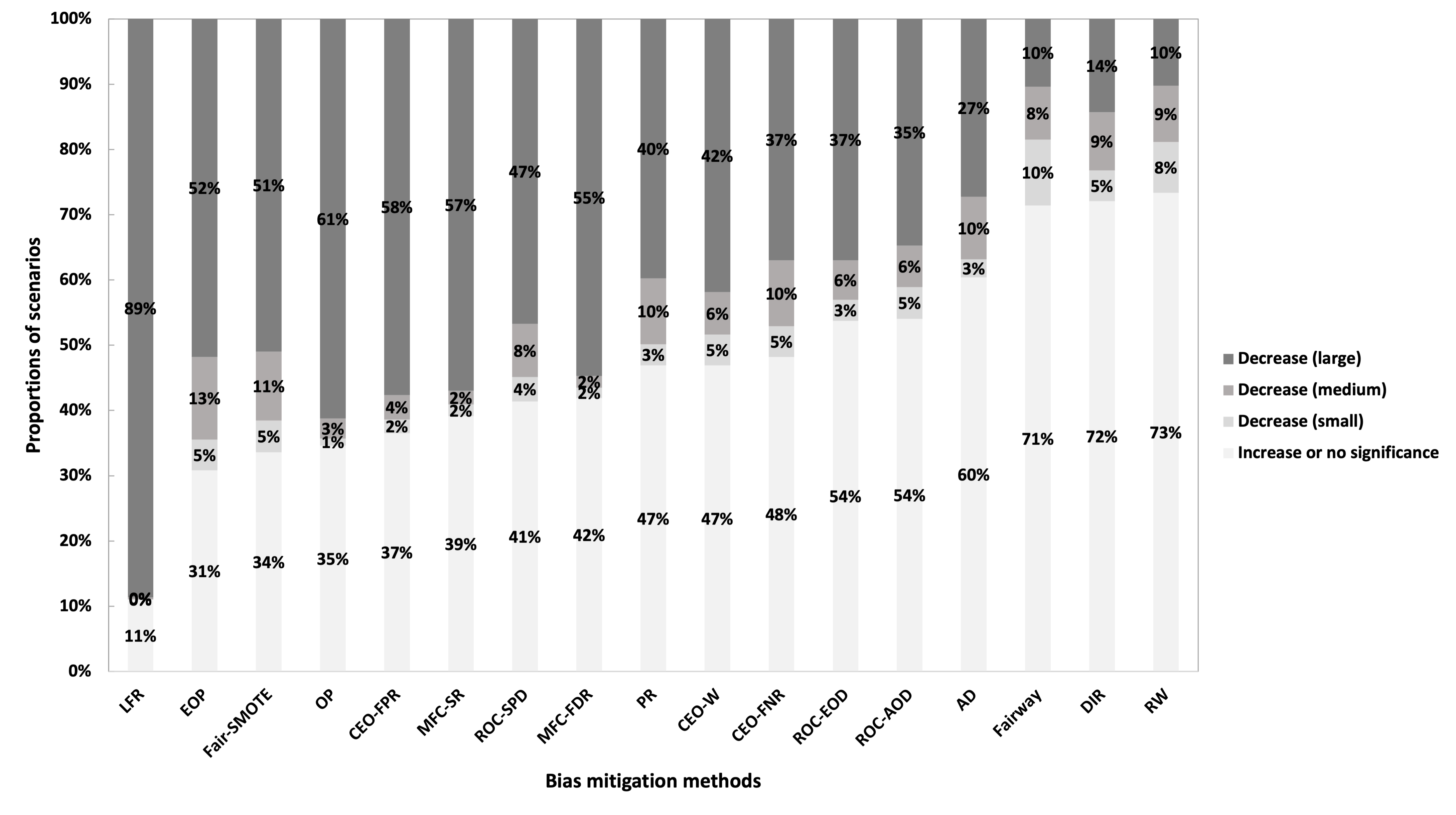}
\caption{\textbf{RQ1.2:} Effect distribution of different bias mitigation methods in ML performance. We find that 6 out of 17 bias mitigation methods significantly decrease ML performance in less than 50\% of scenarios.}\label{fig:methodeffect_per}
\end{center}
\end{figure}

\begin{table*}[!tp]
\centering
\tiny
\caption{\textbf{RQ1.2:} Mean rank of each bias mitigation method in terms of different ML performance metrics. For each metric, we highlight the top-ranked bias mitigation method (i.e., the method with the smallest rank value) in shading. For any ML performance metric, none of the 17 bias mitigation methods can consistently achieve better results than other methods in all the task-model pairs.}
\label{rankresult_p}
\begin{tabular}{l|rrrr|rrrr|rrrrrrr|rr}
\hline
Metric & OP & LFR & RW & DIR & PR & AD & {\tabincell{c}{MFC\\-FDR}} & {\tabincell{c}{MFC\\-SR}} & {\tabincell{c}{ROC\\-SPD}} & {\tabincell{c}{ROC\\-AOD}} & {\tabincell{c}{ROC\\-EOD}} & {\tabincell{c}{CEO\\-FNR}} & {\tabincell{c}{CEO\\-FPR}} & {\tabincell{c}{CEO\\-W}} & EOP & {\tabincell{c}{Fair\\-way}} & {\tabincell{c}{Fair\\-SMOTE}}\\
\hline
F-P & 12 & 16 & 6 & 7 & \cellcolor{gray!25}5 & 6 & 12 & 11 & 7 & 6 & 7 & 9 & 8 & 7 & 9 & 6 & 8\\
F-R & 8 & 9 & 9 & \cellcolor{gray!25}7 & 9 & 10 & \cellcolor{gray!25}7 & \cellcolor{gray!25}7 & 10 & 8 & 8 & \cellcolor{gray!25}7 & 9 & 9 & 9 & \cellcolor{gray!25}7 & 10\\
F-F1 & 9 & 11 & 7 & 5 & 9 & 8 & 10 & 12 & 10 & 8 & 7 & 6 & 9 & 9 & 9 & \cellcolor{gray!25}4 & 9\\
UnF-P & 11 & 16 & 8 & 6 & 10 & 8 & 6 & 6 & 9 & 7 & 7 & 7 & 10 & 9 & 9 & \cellcolor{gray!25}5 & 9\\
UnF-R & 11 & 9 & \cellcolor{gray!25}7 & 9 & \cellcolor{gray!25}7 & \cellcolor{gray!25}7 & 10 & 9 & \cellcolor{gray!25}7 & 8 & 8 & 10 & 8 & \cellcolor{gray!25}7 & 9 & 8 & 8\\
UnF-F1 & 10 & 14 & 6 & 7 & 8 & 6 & 11 & 9 & 8 & 7 & 8 & 9 & 10 & 8 & 9 & \cellcolor{gray!25}5 & 8\\
Acc & 10 & 14 & 4 & 3 & 7 & 5 & 14 & 13 & 11 & 10 & 11 & 6 & 8 & 6 & 9 & \cellcolor{gray!25}2 & 9\\
Mac-P & 11 & 16 & 5 & 4 & 5 & 5 & 13 & 12 & 10 & 9 & 10 & 7 & 7 & 6 & 10 & \cellcolor{gray!25}3 & 10\\
Mac-R & 11 & 16 & 7 & 6 & 11 & 8 & 8 & 9 & 6 & 3 & \cellcolor{gray!25}2 & 10 & 12 & 11 & 9 & 6 & 8\\
Mac-F1 & 11 & 16 & 6 & 5 & 10 & 6 & 11 & 12 & 8 & 5 & 5 & 9 & 11 & 9 & 9 & \cellcolor{gray!25}3 & 8\\
MCC & 11 & 16 & 6 & 5 & 10 & 7 & 10 & 11 & 8 & 4 & \cellcolor{gray!25}3 & 9 & 11 & 9 & 10 & \cellcolor{gray!25}3 & 8\\
\hline
\end{tabular}
\end{table*}

\subsubsection{RQ1.2: ML Performance Comparison among Methods}\label{perform_12}
To compare different bias mitigation methods in terms of ML performance degradation, we first calculate the proportions of scenarios, i.e., \emph{(task-model pair, ML performance metric)} combinations, that fall into different effects after applying each bias mitigation method. Fig. \ref{fig:methodeffect_per} show the results. We present the bias mitigation methods in descending order by the proportion of scenarios with significantly decreased ML performance.
Among the bias mitigation methods that we study, RW performs the best in retaining ML performance, while LFR performs the worst. Specifically, RW significantly decreases ML performance in the fewest (27\%) scenarios, with a large effect in 10\%. The good performance of RW is because it adjusts only the weights of examples in training data and does not modify their features or labels, avoiding introducing additional noise.
In contrast, LFR learns data representations that obfuscate information about protected attributes, and decreases ML performance significantly with a large effect in 89\% of scenarios, 79\% more than RW. Given the big difference in ML performance of different methods, researchers and practitioners should select their bias mitigation method carefully with ML performance considered. Our comprehensive results provide implications for researchers and practitioners to choose appropriate bias mitigation methods according to different ML performance requirements.

Methods that mitigate bias with ML performance taking into account (e.g., DIR, Fairway, and AD) tend to better retain ML performance. Many of existing methods mitigate bias without considering ML performance, which causes the significant decrease in ML performance. We take LFR, EOP, and Fair-SMOTE as examples, which significantly decrease ML performance in the most scenarios as shown Fig. \ref{fig:methodeffect_per}. They all do not consider ML performance during bias mitigation. Specifically, LFR obfuscates information about protected attributes to mitigate bias; EOP changes output labels to optimize equalized odds; Fair-SMOTE generates new data points to balance samples in different groups to reduce data bias.
% OP modifies data features and labels to pursue group fairness. As a result, the loss of information and modification of ground truth data decrease ML performance much. 
In contrast, DIR improves fairness while preserving rank-ordering within groups; AD maximizes accuracy and reduces evidence of protected attributes simultaneously; Fairway employs multi-objective optimization to maximize model performance while making it fair. As a result, they minimally damage ML performance (as shown in Fig. \ref{fig:methodeffect_per}).

Furthermore, for each ML performance metric, we follow previous work \cite{biswas2020machine} to compute the average rank of each bias mitigation method in the 56 task-model pairs. The smaller the rank value, the less the corresponding method reduces the ML performance metric value. Table \ref{rankresult_p} shows the results and highlights the top-ranked method for each metric. We observe that the performance drop can be significantly different across different performance metrics.  For example, ROC-EOD ranks 2nd for Mac-R among the 17 methods, but it ranks 11th for Acc. This is because different ML performance metrics do not necessarily have a positive correlation and many of them are even negatively correlated (as shown in Table 2). In this case, a bias mitigation method cannot retain its ML performance regarding every metric. Considering the diverse ML performance requirements in real-world applications, this finding suggests researchers take comprehensive metrics for bias mitigation research so as to capture any decrease in performance caused by bias mitigation methods, which further evidences the motivation of our comprehensive study.

\finding{\textbf{Finding 2:} Among the 17 bias mitigation methods that we study, RW performs the best in retaining ML performance, while LFR performs the worst. Methods that mitigate bias with ML performance taking into account (e.g., Fairway, DIR, and AD) tend to better retain ML performance. The performance drop can be significantly different across different performance metrics. }

\subsection{RQ2: Influence on Fairness}\label{fairresult}
This section analyzes the evaluation results for four fairness metrics by answering two specific questions:

\textbf{RQ2.1 (Effects on fairness metric values):} \emph{How do the values of fairness metrics change after bias mitigation?} First, we investigate whether the values of fairness metrics are significantly changed after we apply bias mitigation methods, and whether the changes in different fairness metrics are significantly correlated.

\textbf{RQ2.2 (Fairness comparison among methods):} \emph{How do different bias mitigation methods affect ML software fairness?} Second, we compare existing bias mitigation methods in terms of different fairness metrics. The results provide implications for the choice of bias mitigation methods in application scenarios where fairness is critical.

\begin{figure}[!t]
\begin{center}
\includegraphics[width=0.6\columnwidth]{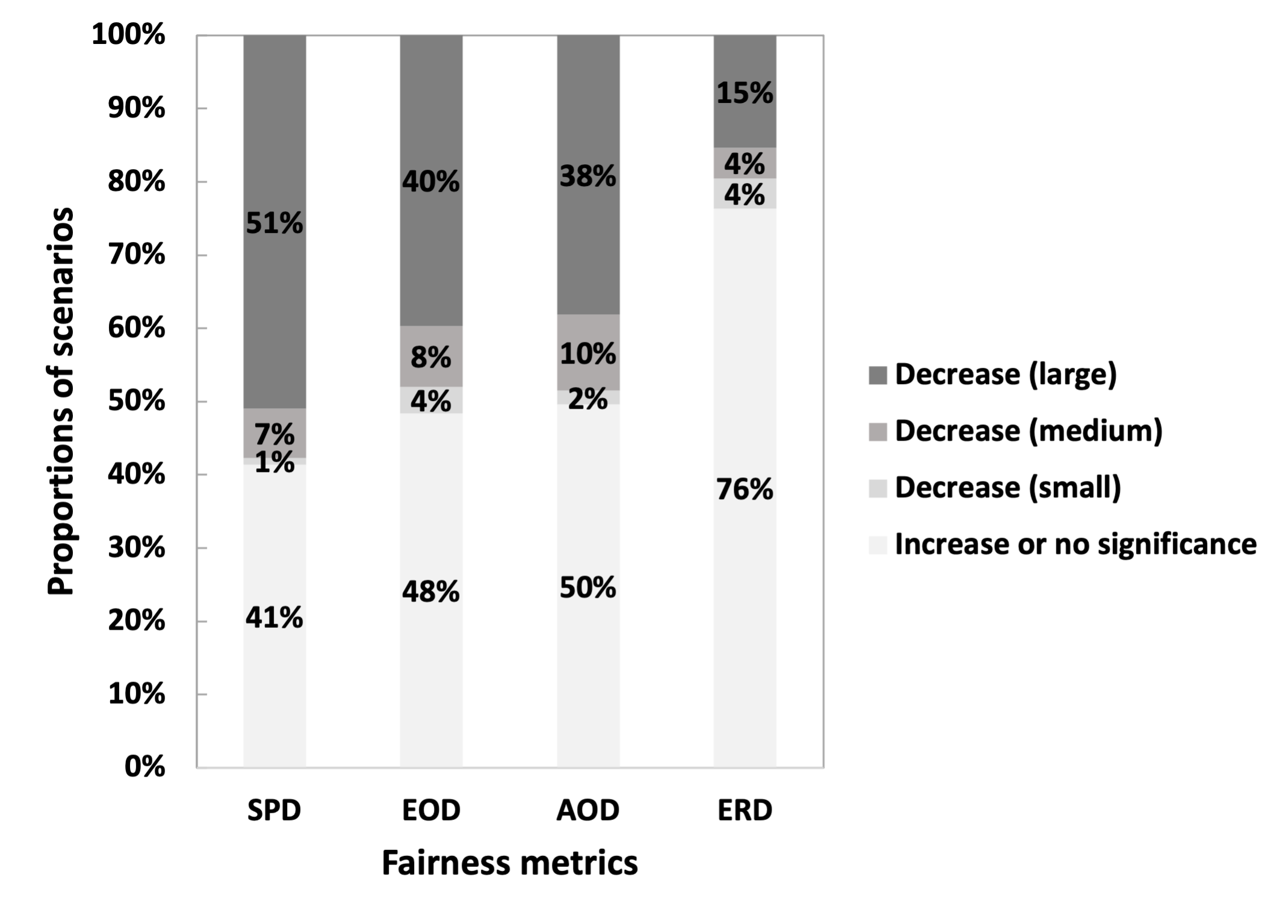}
\caption{\textbf{RQ2.1:} Effects of bias mitigation methods on different fairness metrics. After applying existing bias mitigation methods, the values of the four fairness metrics significantly decrease (i.e., fairness is improved) in an average of 46\% of applications.}\label{fig:effect_fair}
\end{center}
\end{figure}

\subsubsection{RQ2.1: Effects on Fairness Metric Values}
We follow the procedures in Section \ref{performance_effect} to analyze the value changes of the four fairness metrics caused by bias mitigation methods, and present the results in Fig. \ref{fig:effect_fair}. We observe that after applying existing bias mitigation methods, the values of the four fairness metrics (i.e., SPD, EOD, AOD, and ERD) significantly decrease (i.e., fairness is improved) in 59\%, 52\%, 50\%, and 24\% of applications, with an average of 46\%. Considering that the main goal of bias mitigation methods is to improve software fairness (i.e, decrease the values of fairness metrics), our results warn the research community about the limitations of existing bias mitigation methods, especially in reducing ERD.

Furthermore, similar to Section \ref{performance_effect}, we calculate the correlation between each two fairness metrics in terms of their value differences before and after bias mitigation. Table \ref{corrresult_f} shows the results. We observe that the correlation coefficient between AOD and EOD is 0.906 at a significant level, and the significantly positive correlation holds in all the 56 task-model pairs. This suggests that bias mitigation methods that perform well in AOD may also perform well in EOD, and that the evaluation results of bias mitigation methods using AOD may also provide guidance for method selection in application scenarios that pursue a low EOD. Therefore, it is not surprising to find that some previous work \cite{sigsoftHortZSH21} employ only one of them for evaluation. 
In addition, SPD and AOD have a significantly positive correlation in 55 out of the 56 task-model pairs, and the overall correlation coefficient between them is 0.865.

In other fairness metric pairs, we find that the correlations vary in different task-model pairs, which is consistent with the finding in previous work \cite{biswas2020machine} that analyzes the metric correlation in only two bias mitigation tasks. For instance, we find that SPD and EOD have a significantly positive correlation in 34 out of the 56 task-model pairs.  

Additionally, we observe that ERD has a negative correlation with other fairness metrics overall. This indicates that bias mitigation methods that decrease the ERD value often increase bias in terms of other three fairness metrics. Moreover, ERD changes do not have a consistent correlation with changes of any other metric across the task-model pairs. This may be attributed to the basis of these fairness metrics. Specifically, SPD, EOD, and AOD are calculated based on the positives (i.e., all positives for SPD and AOD, and true positives for EOD), while ERD relies on false negatives and false positives. This indicates that other metrics may not act as a proxy of ERD. However, some existing work \cite{jieMLsurvey,sigsoftHortZSH21,fairsmotepaper,fairwaypaper} does not take ERD into account. To perform a comprehensive evaluation, we suggest that researchers follow Biswas and Rajan \cite{biswas2020machine,sigsoftBiswasR21} to consider this metric in future work.

\finding{\textbf{Finding 3:} Existing bias mitigation methods improve fairness significantly (in terms of SPD, AOD, EOD, and ERD) in an average of 46\% of the studied applications. In particular, existing methods significantly improve fairness measured by ERD in 24\% of scenarios. Additionally, 
fairness improvement measured by different metrics are not necessarily correlated. In particular, ERD changes do not have a consistent correlation with the changes of any other fairness metric.}

\begin{table}[!tp]
\centering
\small
\caption{\textbf{RQ2.1:} Correlation between fairness metrics. In the table, * indicates a significant correlation ($p$-value $\textless$ 0.05) overall, and numbers in parentheses indicate the number of task-model pairs the correlation shares the same pattern with the overall correlation. We find that AOD and EOD have a consistently positive correlation across all the task-model pairs, and that ERD does not have a consistent correlation with any other metric.}
\label{corrresult_f}
\begin{tabular}{l|rrr}
\hline
  & AOD & EOD& ERD\\
 \hline
SPD& 0.865*(55/56)& 0.688*(34/56)& -0.154*(10/56)\\
AOD& -& 0.906*(56/56)& -0.218*(8/56)\\
EOD& -& -& -0.263*(16/56)\\
\hline
\end{tabular}
\end{table}

\begin{figure}[!t]
\begin{center}
\includegraphics[width=1\columnwidth]{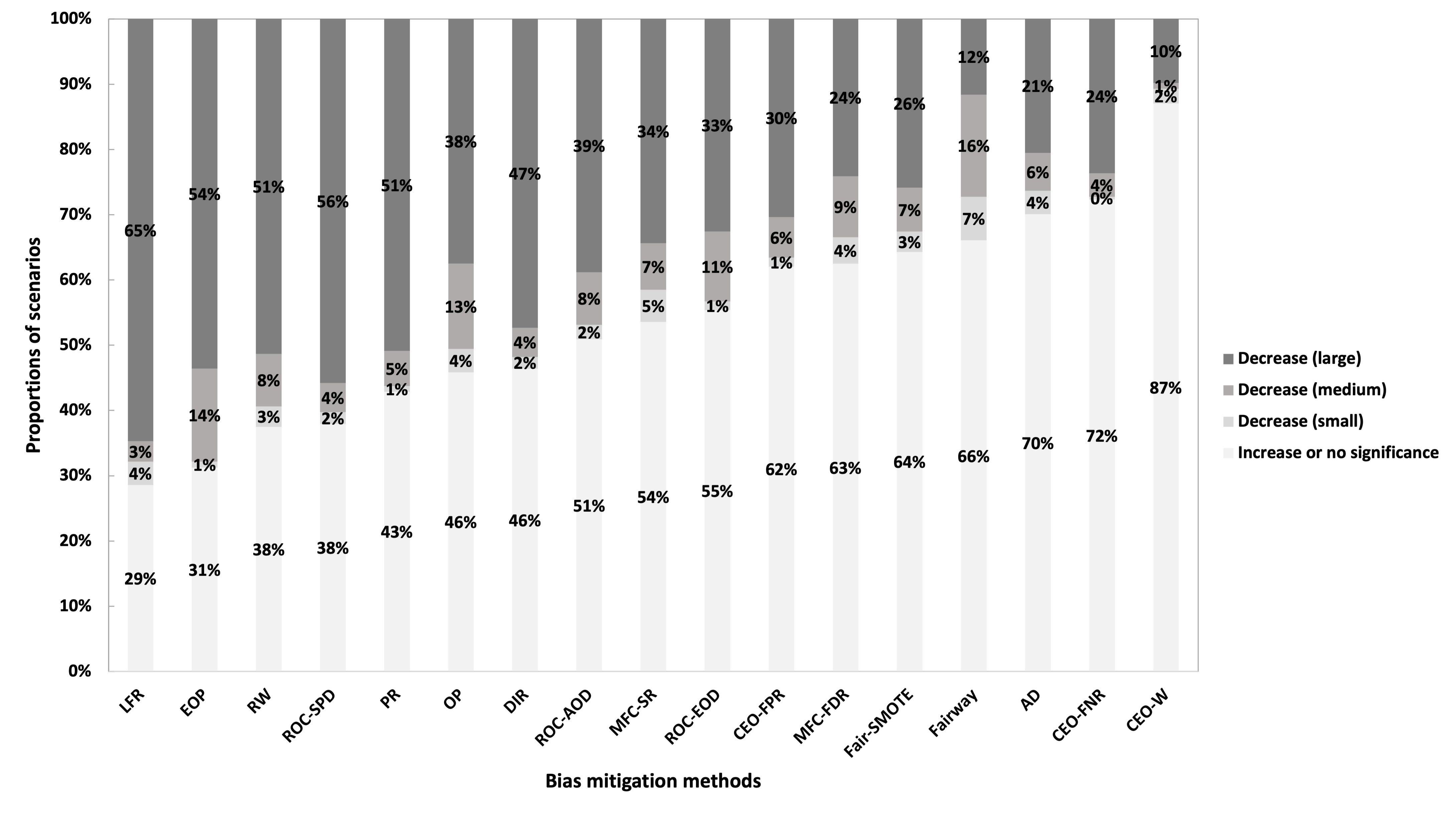}
\caption{\textbf{RQ2.2:} Effect distribution of different bias mitigation methods in fairness. We observe that 7 out of 17 bias mitigation methods significantly reduce bias in more than 50\% of scenarios.}\label{fig:methodeffect_fair}
\end{center}
\end{figure}

\subsubsection{RQ2.2: Fairness Comparison among Methods}
Similar to Section \ref{perform_12}, we calculate the proportions of scenarios, i.e., \emph{(task-model pair, fairness metric)} combinations, that fall into different effects after applying each bias mitigation method. Fig. \ref{fig:methodeffect_fair} shows the results. We present bias mitigation methods in descending order by the proportion of scenarios with significantly improved fairness. We find that LFR significantly improves fairness in the most scenarios (71\%) among the 17 bias mitigation methods. As described in Section 4.1.2, it also significantly decreases ML performance in the most scenarios, further providing empirical evidence for the existence of fairness-performance trade-off. In contrast, CEO-W significantly improves fairness in the fewest (23\%) scenarios, 48\% fewer than LFR. Moreover, although fairness improvement is the main goal of bias mitigation methods, only 7 out of the 17 bias mitigation methods (i.e., LFR, EOP, RW, ROC-SPD, PR, OP, and DIR) can significantly improve fairness in more than 50\% of scenarios. Therefore, a community effort is required to bring software fairness improvement to a level where it becomes more effective and usable in practice.

Methods designed for optimizing specific metrics (e.g., CEO-W, CEO-FNR, MFC-FDR, and CEO-FPR) tend to have poor overall fairness. For example, CEO-FNR aims at minimizing the false negative rate difference across population groups, and thus the ML model is optimized for this goal, causing damage to other goals. In contrast, the LFR method, which reduces the information of protected attributes rather than optimizing for a specific fairness goal, achieves the best fairness over all the metrics.

\begin{table*}[!tp]
\centering
\tiny
\caption{\textbf{RQ2.2:} Mean rank of each bias mitigation method in terms of different fairness metrics. For each metric, we highlight the top-ranked bias mitigation method (i.e., the method with the smallest rank value) in shading. It is difficult for bias mitigation methods to achieve fairness with respect to all the metrics.}
\label{rankresult_f}
\begin{tabular}{l|rrrr|rrrr|rrrrrrr|rr}
\hline
Metric & OP & LFR & RW & DIR & PR & AD & {\tabincell{c}{MFC\\-FDR}} & {\tabincell{c}{MFC\\-SR}} & {\tabincell{c}{ROC\\-SPD}} & {\tabincell{c}{ROC\\-AOD}} & {\tabincell{c}{ROC\\-EOD}} & {\tabincell{c}{CEO\\-FNR}} & {\tabincell{c}{CEO\\-FPR}} & {\tabincell{c}{CEO\\-W}} & EOP & {\tabincell{c}{Fair\\-way}} & {\tabincell{c}{Fair\\-SMOTE}}\\
\hline
SPD & 9 & \cellcolor{gray!25}2 & 5 & 8 & 6 & 10 & 10 & 7 & 4 & 10 & 11 & 11 & 11 & 14 & 6 & 10 & 10\\
AOD & 9 & \cellcolor{gray!25}2 & 5 & 6 & 8 & 13 & 8 & 7 & 7 & 8 & 9 & 10 & 12 & 15 & 4 & 9 & 10\\
EOD & 9 & \cellcolor{gray!25}3 & 7 & 7 & 8 & 13 & 7 & 4 & 8 & 6 & 7 & 11 & 12 & 14 & 5 & 10 & 10\\
ERD & 8 & 15 & 7 & 8 & 13 & 8 & 13 & 11 & \cellcolor{gray!25}5 & 6 & 8 & 8 & 8 & \cellcolor{gray!25}5 & \cellcolor{gray!25}5 & 7 & 8\\
\hline
\end{tabular}
\end{table*}

Next, we calculate the mean rank of each bias mitigation method in the 56 task-model pairs for each fairness metric. Table \ref{rankresult_f} shows the results. Different fairness metrics yield significantly different ranking results of the bias mitigation effectiveness. For example, LFR is the top-ranked method for SPD, but it ranks 15th among the 17 methods for ERD. It is because that SPD often has a negative correlation with ERD (as shown in Table 4). This further evidences that it is difficult for bias mitigation methods to achieve fairness with respect to all the metrics \cite{biswas2020machine,berk2021fairness,chouldechova2017fair}. It suggests that besides the fairness-performance trade-off, researchers also need to take into account the trade-off between different fairness metrics when designing bias mitigation methods, especially considering the diverse fairness requirements in real-world applications.

\finding{\textbf{Finding 4:} LFR significantly improves fairness in the most scenarios (71\%) among the 17 bias mitigation methods that we study. Moreover, only 7 of the 17 methods significantly improve fairness in more than half of the scenarios. In particular, methods designed for optimizing specific metrics (e.g., CEO-W, CEO-FNR, MFC-FDR, and CEO-FPR) tend to have poor overall fairness. Different fairness metrics yield significantly different ranking results of the bias mitigation effectiveness. For example, LFR is the top-ranked method for SPD, but it ranks 15th among the 17 methods for ERD. }

\subsection{RQ3: Fairness-performance Trade-off}\label{traderesult}
In this section, we present the measurement results of 20 types of fairness-performance trade-offs, i.e., combinations of four fairness metrics (SPD, AOD, EOD, and ERD) and five ML performance metrics (Acc, Mac-P, Mac-R, Mac-F1, and MCC), achieved by different bias mitigation methods using Fairea \cite{sigsoftHortZSH21}.

\begin{figure}[!t]
\begin{center}
\includegraphics[width=1\columnwidth]{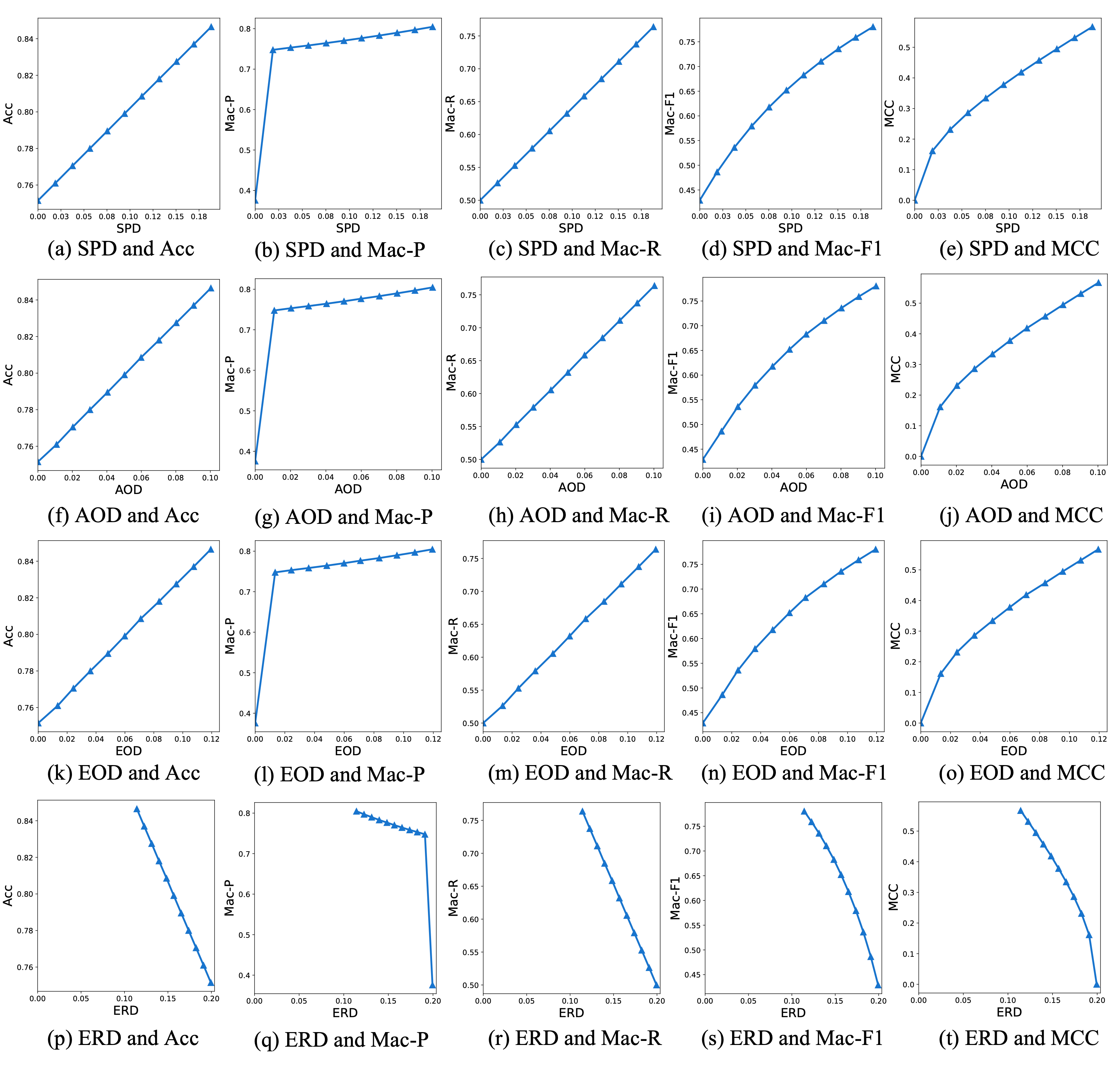}
\caption{\textbf{RQ3:} Fairness-performance trade-off baselines for LR model in Adult-Sex task. ERD shows a different trade-off pattern from other fairness metrics in the baselines constructed by Fairea.}\label{fig:examplebase}
\end{center}
\end{figure}

As described in Section \ref{faireades}, the first step of Fairea is to construct the trade-off baseline using a series of pseudo models generated via model behavior mutation. For each \emph{(task-model pair, fairness-performance metric pair)} combination,  we construct the baseline separately. As a result, we construct a total of $8\times7\times4\times5 = 1,120$ baselines.
Based on the 1,120 baselines, we observe that the pseudo models show a fairness-performance trade-off for SPD, AOD, and EOD, i.e., the higher the value of SPD, AOD, or EOD, the higher the value of each ML performance metric. 
However, we fail to construct such trade-off baselines for ERD. We take the baselines constructed for the LR model in the Adult-Sex task (shown in Fig. \ref{fig:examplebase}) as an example. Based on the generated pseudo models, ERD shows a different trade-off pattern from SPD, AOD, and EOD. 
The different trade-off patterns shown by ERD and other fairness metrics can be explained by their different definitions (i.e., calculation methods). Based on their calculation methods presented in Section \ref{fairnessmetrics}, we can find that when all samples are predicted as the same label (i.e., the model achieves the worst ML performance), SPD, AOD, and EOD all equal to 0 (their minimum), but ERD equals to the difference of favorable rates between the privileged and unprivileged groups (not its minimum). This means that ERD does not meet the hypothesis behind Fairea.
The baseline in Fairea is constructed based on the hypothesis that the fairness of an ML model can be improved by sacrificing its performance with the increased model behavior mutation degrees. However, from Fig. \ref{fig:examplebase}(p)$\sim$(t), we observe that the increased mutation degrees not only result in the sacrifice of ML performance, but also cause the increased ERD value (i.e., decreased fairness in terms of ERD). 

Since Fairea fails to construct the trade-off baseline for ERD, the effectiveness region division method and the trade-off quantification method provided by Fairea are also not applicable to ERD. Therefore, we consider 15 types of fairness-performance trade-offs, i.e., combinations of three fairness metrics (SPD, AOD, and EOD) and five ML performance metrics (Acc, Mac-P, Mac-R, Mac-F1, and MCC), and the corresponding $8\times7\times3\times5 = 840$ baselines, in the rest of the paper.

Based on the measurement results of the 15 trade-offs, we explore RQ3 by answering two specific questions:

\textbf{RQ3.1 (Effectiveness region distribution):} \emph{What effectiveness regions do existing bias mitigation methods fall into according to Fairea?} This research question evaluates the overall effectiveness of existing bias mitigation methods by analyzing how the mitigation cases achieved by them are matched into the five effectiveness regions in Fig. \ref{fig:fairea}(b).

\textbf{RQ3.2 (Quantitative assessment of trade-off):} \emph{What fairness-performance trade-off do existing bias mitigation methods achieved according to the quantitative assessment of Fairea?} This research question evaluates existing bias mitigation methods in terms of quantitative assessment of fairness-performance trade-off. Specifically, we calculate the area shown in Fig. \ref{fig:fairea}(a) for each bias mitigation method and then compare these methods quantitatively.

\begin{figure} 
    \centering
       \includegraphics[width=1\linewidth]{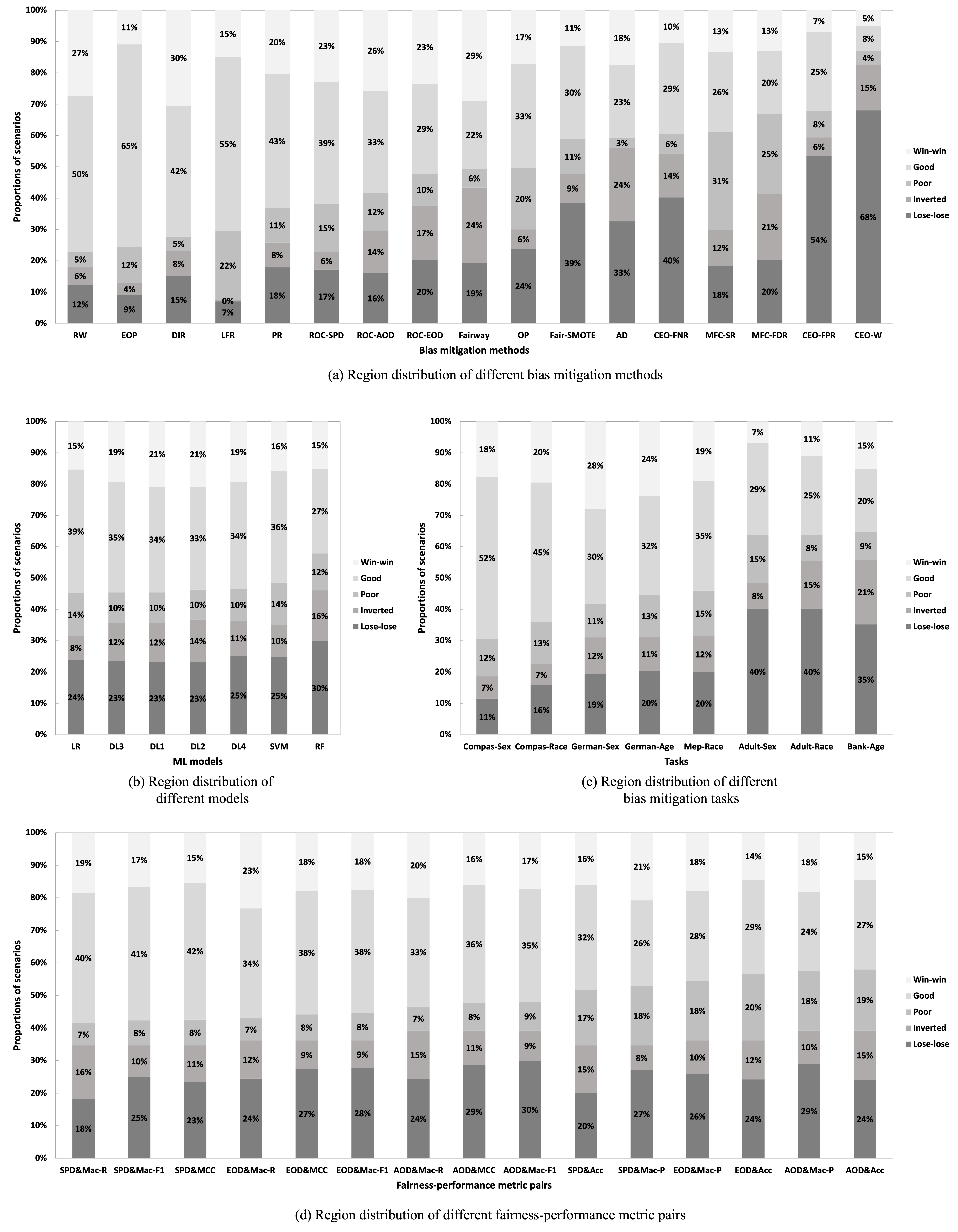}
  \caption{\textbf{RQ3.1:} Proportion of mitigation cases that fall into each mitigation region organized by (a) bias mitigation methods, (b) models, (c) bias mitigation tasks, and (d) fairness-performance metric pairs. We observe that a notable proportion (25\%) of mitigation cases fall into a lose-lose trade-off region, and that the effectiveness of bias mitigation methods depends on models, tasks, the selection of protected attributes, and the set of metrics used to assess fairness and ML performance.}
  \label{fig:regiondis} 
\end{figure}

\subsubsection{RQ3.1: Effectiveness Region Distribution}\label{sectionregion}
We use the 840 baselines to evaluate the effectiveness of 17 bias mitigation methods in 56 task-model pairs. We apply each bias mitigation method to each task-model pair 50 times and treat each individual run as a mitigation case. Therefore, we have $8\times7\times50 = 2,800$ mitigation cases for each bias mitigation method.\footnote{For the OP method, we have $6\times7\times50 = 2,100$ mitigation cases, as we apply it in only six tasks.} 
In Fig. \ref{fig:regiondis}(a), we present the overall results of different bias mitigation methods for all the task-model pairs and fairness-performance metric pairs.

From Fig. \ref{fig:regiondis}(a), we find that RW achieves a win-win or good trade-off in the most scenarios (77\%), while CEO-W does so in the fewest scenarios (13\%). From this perspective, RW makes the best trade-off in general, achieving a win-win trade-off in 27\% (ranging from 23\% to 34\% according to different fairness-performance metric pairs) and a good trade-off in 50\% of cases (ranging from 38\% to 63\%). A possible reason behind the superiority of RW is that it modifies the weights of different training samples so that it tackles the training data imbalance problem, which has been recognized as a root cause of bias in ML software \cite{fairsmotepaper,csurMehrabiMSLG21}. Meanwhile, it does not modify features or labels of training data, which avoids introducing additional noise and thus retains ML performance. Nevertheless, we still observe that 12\% of mitigation cases fall into a lose-lose trade-off after applying RW, and the situation is even worse for other bias mitigation methods. For example, CEO-W achieves a lose-lose trade-off in 68\% of cases. Overall, the proportion of cases that fall into the lose-lose region obtained by all the studied methods is 25\%. The notable proportion of the lose-lose trade-off is also observed in previous work \cite{sigsoftHortZSH21}, and our results increase the confidence of this finding with more bias mitigation methods and fairness-performance metric pairs. One possible reason behind the high percentage of the lose-lose trade-off is that bias mitigation methods are often designed to optimize one fairness metric, which may affect other fairness metrics. For instance, ROC-AOD achieves a lose-lose trade-off between AOD and Acc in 17.3\% of mitigation cases, while achieves a lose-lose trade-off between SPD and Acc in 32.0\% of mitigation cases; ROC-SPD achieves a lose-lose trade-off between SPD and Acc in 1.3\% of mitigation cases, while achieves a lose-lose trade-off between AOD and Acc in 22.6\% of mitigation cases.

Furthermore, we organize the results by different models, bias mitigation tasks, and fairness-performance metric pairs, to analyze whether the region distribution is influenced by these factors. We find that the effectiveness of bias mitigation methods depends on models, tasks, the selection of protected attributes, and the set of metrics used to assess fairness and ML performance. It is because the different characteristics of different models, tasks, protected attributes, and metrics pose different degrees of difficulty for bias mitigation. The results are presented in Figs. \ref{fig:regiondis}(b), (c), and (d), respectively, and the analysis is as follows. 

\emph{Comparison among models:} From Fig. \ref{fig:regiondis}(b), we observe that the region distribution is model-dependent. Although LR, SVM, RF, DL1, DL2, DL3, and DL4 share similar proportions of mitigation cases that fall into the lose-lose, poor, and win-win trade-off regions, they have obvious differences in the proportions of the inverted and good trade-offs. Specifically, LR, SVM, DL1, DL2, DL3, and DL4 achieve a relatively higher percentage of the good trade-off (a lower percentage of the inverted trade-off) than RF. For example, the difference between LR and RF in the proportion of the good trade-off is 12\% (i.e., 39\% v.s. 27\%). Therefore, we suggest that researchers consider different models when evaluating the effectiveness of bias mitigation methods.

\emph{Comparison among tasks:} From Fig. \ref{fig:regiondis}(c), we find that the region distribution is task-dependent. Existing methods achieve worse fairness-performance trade-off on imbalanced datasets than on balanced datasets. For example, the Compas dataset is the most balanced among all the datasets, and thus the classification on it is easier compared to other tasks. Therefore, as shown in Fig. 8(c), we can observe that existing bias mitigation methods can retain ML performance well while mitigating bias (i.e., having a good fairness-performance trade-off). In contrast, existing methods achieve the worst trade-off in the most imbalanced dataset (i.e., the Bank dataset), where the major label accounts for 82.8\%.
Furthermore, we observe that even for the same dataset, the selection of different protected attributes also affects the region distribution. For example, the proportion of the poor trade-off in the Adult-Sex is 15\%, nearly twice the corresponding proportion in Adult-Race task (8\%). This finding suggests that (1) researchers evaluate bias mitigation methods in diverse tasks to improve the generalizability of the results, (2) practitioners need to be careful when choosing bias mitigation methods for their own tasks based on existing evaluation results on other tasks, and (3) practitioners may need to choose different bias mitigation methods for different protected attributes of the same dataset.

\emph{Comparison among fairness-performance metric pairs:} From Fig. \ref{fig:regiondis}(d), we find that different fairness-performance metric pairs result in different region distributions. In particular, the mitigation cases that obtain a win-win or good trade-off between AOD and Acc account for 42\%, the lowest among all the fairness-performance metric pairs. In contrast, the corresponding proportion between SPD and Mac-R is 12\% higher (i.e., 59\%). This finding further demonstrates the necessity of this study that measures the trade-off in terms of various fairness-performance metric pairs to obtain more general and comprehensive results.

\finding{\textbf{Finding 5:} According to the trade-off classification framework provided by Fairea, RW has the best fairness-performance trade-off (with a win-win or good trade-off in 77\% of the scenarios). Furthermore, on average, existing methods damage both fairness and ML performance (i.e., a lose-lose trade-off) in 25\% of the studied scenarios. The effectiveness of bias mitigation methods depends on models, tasks, the selection of protected attributes, and the set of metrics used to assess fairness and ML performance. In particular, existing methods achieve worse fairness-performance trade-off on imbalanced datasets than on balanced datasets.}

\begin{table*}[!tp]
\setlength\tabcolsep{4pt}
\centering
\tiny
\caption{\textbf{RQ3.2:} Numbers of scenarios where each method achieves the best quantitative result, organized by different bias mitigation tasks. For each task, we highlight the bias mitigation method that achieves the best trade-off in the most scenarios in shading. Overall, none of the existing bias mitigation methods can achieve a better trade-off than other methods in all the scenarios, as even the best method that we find outperforms other methods in 30\% of scenarios.}
\label{areatable}
\begin{tabular}{l|rrrr|rrrr|rrrrrrr|rr}
\hline
Task & OP & LFR & RW & DIR & PR & AD & {\tabincell{c}{MFC\\-FDR}} & {\tabincell{c}{MFC\\-SR}} & {\tabincell{c}{ROC\\-SPD}} & {\tabincell{c}{ROC\\-AOD}} & {\tabincell{c}{ROC\\-EOD}} & {\tabincell{c}{CEO\\-FNR}} & {\tabincell{c}{CEO\\-FPR}} & {\tabincell{c}{CEO\\-W}} & EOP & {\tabincell{c}{Fair\\-way}} & {\tabincell{c}{Fair\\-SMOTE}}\\
\hline
Adult-Sex & 0\% & 0\% & 0\% & 11\% & \cellcolor{gray!25}20\% & 15\% & 0\% & 0\% & 6\% & \cellcolor{gray!25}20\% & 4\% & 3\% & 0\% & 0\% & 13\% & 15\% & 14\%\\
Adult-Race & 0\% & 0\% & 7\% & \cellcolor{gray!25}46\% & 0\% & 23\% & 5\% & 2\% & 17\% & 15\% & 6\% & 0\% & 0\% & 0\% & 4\% & 9\% & 7\%\\
Compas-Sex & \cellcolor{gray!25}77\% & 0\% & 4\% & 50\% & 14\% & 0\% & 0\% & 0\% & 3\% & 0\% & 0\% & 0\% & 0\% & 0\% & 0\% & 20\% & 0\%\\
Compas-Race & 48\% & 8\% & 13\% & 34\% & 54\% & 63\% & 0\% & 0\% & 3\% & 8\% & 0\% & 0\% & 0\% & 0\% & 0\% & \cellcolor{gray!25}66\% & 0\%\\
German-Sex & 0\% & 34\% & 0\% & 32\% & 17\% & 0\% & 41\% & \cellcolor{gray!25}63\% & 36\% & 51\% & 50\% & 0\% & 5\% & 5\% & 0\% & 61\% & 7\%\\
German-Age & 0\% & 34\% & 28\% & 34\% & 0\% & 0\% & 0\% & \cellcolor{gray!25}67\% & 28\% & 48\% & 45\% & 0\% & 11\% & 0\% & 0\% & 25\% & 8\%\\
Bank-Age & - & 3\% & 43\% & 0\% & 7\% & 0\% & 2\% & 7\% & 17\% & 5\% & 15\% & \cellcolor{gray!25}57\% & 0\% & 0\% & 15\% & 22\% & 2\%\\
Mep-Race & - & 0\% & 9\% & 14\% & 32\% & 28\% & 16\% & 0\% & 24\% & \cellcolor{gray!25}39\% & 36\% & 0\% & 0\% & 0\% & 0\% & 18\% & 8\%\\
\hline
Overall & 20\% & 10\% & 13\% & 27\% & 18\% & 16\% & 8\% & 17\% & 17\% & 23\% & 19\% & 7\% & 2\% & 0\% & 4\% & \cellcolor{gray!25}30\% & 6\%\\
\hline
\end{tabular}
\end{table*}

\subsubsection{RQ3.2: Quantitative Assessment of Trade-off}
We finally use Fairea to quantify the fairness-performance trade-off of different bias mitigation methods. Fairea quantifies only the cases that fall into the good trade-off region, as the other regions are either dominating the original model (the win-win region), dominated by the Fairea baseline (the poor trade-off region), or do not improve fairness (the inverted and lose-lose trade-off regions) \cite{sigsoftHortZSH21}. 

For each mitigation scenario, i.e., \emph{(task-model pair, fairness-performance metric pair)} combination, we calculate the mean ML performance and fairness results of the 50 runs of each bias mitigation method to indicate its average effectiveness. Then we quantify the trade-off effectiveness of each method with the help of the trade-off baselines. The trade-off quantification enables a direct and convenient comparison among different bias mitigation methods.
We consider win-win > good > poor > inverted > lose-lose trade-off (``>'' means ``better than''), and among good trade-offs, the larger the quantitative result, the better the trade-off achieved by the corresponding bias mitigation method. 
According to this rule, for each bias mitigation task, we calculate the proportion of scenarios where each bias mitigation method achieves the best quantitative result. Table \ref{areatable} shows the result for each task and the overall result in all the tasks. We highlight the highest proportion for each row (i.e., each bias mitigation task). Note that the sum of the numbers in a row is larger than 100\%, since there may be multiple methods to achieve the best quantitative result in a scenario.

At a glance of Table \ref{areatable}, we find that the distributions of the best results vary in different bias mitigation tasks, further demonstrating the aforementioned finding in Section \ref{sectionregion} that the effectiveness of bias mitigation methods is task-dependent. For instance, OP achieves the best quantitative result in 77\% of scenarios in the Compas-Sex task, but the corresponding proportion is 0\% in the Adult-Sex, Adult-Race, German-Sex, and German-Age tasks.

Furthermore, we compare the overall quantitative results achieved by different bias mitigation methods. From this perspective, the top five methods are Fairway (achieving the best quantitative result in 30\% of all the scenarios), DIR (27\%), ROC-AOD (23\%), OP (20\%), and ROC-EOD (19\%).
Overall, even the best method (i.e., Fairway) that we observe outperforms other methods in 30\% of the scenarios. This can be explained by our previous analysis for ML performance and fairness, which demonstrates that no existing methods can consistently have better ML performance or fairness than other methods. This finding suggests practitioners need to choose the bias mitigation method best suited to their intended application scenario(s).

Researchers can also design more generally effective bias mitigation methods. The design of Fairway provides implications for it. Specifically, different from other methods, Fairway combines two different types of bias mitigation methods, i.e., pre-processing and in-processing methods. Similarly, in other software engineering tasks, researchers also find that combining diverse models is able to tackle more ``corner cases'' instances, thereby achieving good results \cite{esemMoussaGS22}. In the future, researchers can propose effective methods that combine bias mitigation strategies across the phases before, in, and after training to achieve better fairness-performance trade-off. In addition, researchers can also design a framework that adaptively provides different appropriate bias mitigation strategies for different scenarios.
\finding{\textbf{Finding 6:} According to the quantitative assessment provided by Fairea, 
the best method that we observe (i.e., Fairway) outperforms other methods in 30\% of the scenarios. There is no bias mitigation method that can achieve the best trade-off in all the scenarios that we study. This indicates that practitioners need to carefully choose the bias mitigation method that best suits their intended application scenario(s).}

\section{Threats to Validity}\label{threat}

\textbf{Threats to construct validity} concern how adequate a concept definition is and how well the indicators represent the concept. The primary threat lies in the definition of ML performance and fairness. To mitigate the threat, we use 11 ML performance metrics and 4 fairness metrics. The 11 ML performance metrics measure not only the performance of individual classes, but also the overall performance; the 4 fairness metrics are the most widely adopted in previous fairness work.

\noindent \textbf{Threats to internal validity} primarily lie in the implementation of the code used in this study. To mitigate the threats, we conduct the experiments based on the widely-adopted IBM AIF360 framework and the released code of Fairway, Fair-SMOTE, and Fairea. We also make all scripts and data publicly available to allow for reproduction and replication.

\noindent \textbf{Threats to external validity} concern the generalizability of our experimental results. To alleviate the threats, we adopt eight bias mitigation tasks, which cover social, financial, and medical domains, and are well adopted in the fairness literature.

Moreover, we employ 17 representative bias mitigation methods proposed in the ML and SE communities. For each method, we use both traditional ML algorithms and DNNs for implementation, reducing the influence of model selection on the generalization of results.

In terms of evaluation measures, we use 11 ML performance metrics, 4 fairness metrics, and 20 types of fairness-performance trade-offs to obtain comprehensive measurement results. Fairea poses a potential threat to the use of fairness-performance trade-off measurements. As discussed in Section \ref{traderesult}, Fairea is constructed based on the hypothesis that the fairness of an ML model can be improved by sacrificing its performance with the increased model behavior mutation degrees. In other words, Fairea cannot be applied to metrics (e.g., ERD) that do not meet the hypothesis, and thus we do not use these metrics in Section \ref{traderesult}.
 
Nevertheless, with increasing attention on software fairness, researchers are proposing more and more bias mitigation methods, fairness metrics, and datasets. In the future, one could replicate this study with more methods, metrics, measures, and datasets.

It is valuable to evaluate existing bias mitigation methods on real deployed software systems, not just libraries as we have done here. Unfortunately it is not possible to use real software in this paper because of the lack of public availability of real software with fairness issues (i.e., lack of open source code and data). This is because decision-making software systems that demand fairness are often human-related and social-critical. For such systems, training data or code is typically unreleased due to privacy and legal concerns. This can also explain the observations of recent surveys \cite{csurMehrabiMSLG21,cscfairness22} that state-of-the-art work on software fairness also conducts their experiments on widely-adopted datasets and ML models like us, rather than on real software. To combat this issue, we use scikit–learn, which is widely used in industry, including big companies such as J.P. Morgan, Amazon, and Microsoft \cite{refer1,refer2}. We also choose the three ML models that are the most widely adopted in the real world \cite{sarker2021machine}. For logistic regression and random forest, an official report about bias in algorithmic decision-making from the UK government \cite{refer4} reveals that when it comes to use cases that demand fairness such as credit scoring decisions, most banks are using simple logistic regression models and random forest models because they have good explainability. For SVM, it has also been widely adopted in critical decision-making applications, such as criminal justice \cite{ghasemi2021application} and credit risk prediction \cite{alabi2020credit}.

In addition, in practice, software engineers may need to mitigate bias regarding multiple protected attributes. However, most of existing bias mitigation methods do not support dealing with multiple protected attributes at the same time. Therefore, we follow previous work \cite{icseZhangH21,sigsoftHortZSH21,fairwaypaper,sigsoftBiswasR21} to consider each protected attribute individually for each bias mitigation task, which may pose a threat to the validity of this study. In the future, when more methods that support this functionality are proposed, one could evaluate them with multiple protected attributes considered at the same time.

\noindent \textbf{Threats to conclusion validity} primarily concern the use of statistical methods. To mitigate the threats, we use the non-parametric Mann Whitney U-test and Spearman's rank correlation, which are widely adopted in SE research \cite{icseArcuriB11,eseBriandEM96}. They suit our purpose well as they do not assume normality. In addition, to improve the reliability of our statistical analysis, we apply each bias mitigation method on different tasks 50 times.

\section{Related Work}\label{related_work}
Fairness has attracted increasing interest from both the ML research  community \cite{csurMehrabiMSLG21} and the SE research community \cite{sigsoftHortZSH21} and also from practitioners as well as researchers. 
For example, Microsoft publishes the ethical principles of AI \cite{microsoftai}, stating that ML software must be fair in real-life applications, and creates a research group named FATE \cite{fategroup} to promote software fairness. 
In addition, researchers,  Brun and Meliou \cite{sigsoftBrunM18} set out a vision for SE research approaches to tackle fairness problems, and a  recent survey on ML testing \cite{jieMLsurvey} classifies fairness as a non-functional software property, surveying SE approaches to tackling it. 

The rising attention on fairness inspires the emergence of a series of fairness testing techniques. Themis \cite{sigsoftGalhotraBM17} generates test suites to measure causal discrimination in software. 
Aeqitas \cite{UdeshiAC18} exploits the inherent robustness property of ML models for directing fairness test generation.
SG \cite{sigsoftAggarwalLNDS19} combines symbolic execution and local explainability to generate test inputs for detecting individual discrimination.
ADF \cite{icseZhangW0D0WDD20} uses gradient computation and clustering to generate individual discriminatory instances for DNNs. Chen et al. \cite{Dabs220710223} provide a comprehensive survey of existing research on fairness testing.

In addition to testing software fairness, researchers also attempt to improve software fairness. Zhang and Harman \cite{icseZhangH21} explore the factors that affect software fairness, and find that enlarging feature set is a possible way to improve fairness. Chakraborty et al. \cite{fairwaypaper} remove ambiguous data points in training data and then apply multi-objective optimization to train fair ML models. To better improve software fairness, the follow-up work of Chakraborty et al. \cite{fairsmotepaper} not only removes ambiguous data points, but also balances the internal distribution of training data. Moreover, IBM launches a software toolkit called AI Fairness 360 (abbreviated as IBM AIF360) \cite{aif360}, which integrates popular fairness improvement methods, including Adversarial Debiasing~\cite{ADVpaper}, Reweighting~\cite{rewpaper}, Reject Option Classification~\cite{ROCpaper}, Learning Fair Representation~\cite{LFRpaper}, etc.

Furthermore, there are some studies that empirically evaluate the effectiveness of different bias mitigation methods. For example, Biswas and Rajan \cite{biswas2020machine,sigsoftBiswasR21} evaluate seven bias mitigation methods on real-world ML models from a crowd-sourced platform and explore the impact of popular pre-processing procedures on ML performance and fairness. Chakraborty et al. \cite{fairwaypaper,fairsmotepaper} compare the bias mitigation methods proposed by them with several methods proposed in the ML community. Hort et al. \cite{sigsoftHortZSH21} propose a model behavior mutation method to quantitatively benchmark and evaluate the fairness-performance trade-off of different bias mitigation methods. However, these evaluations are conducted in terms of limited measurements and methods. In this paper, we present a comprehensive evaluation of existing bias mitigation methods.

\section{Conclusions and Future Work}\label{conclusion}
This paper presents a large-scale empirical study  evaluating 17 representative bias mitigation methods with 11 ML performance metrics, 4 fairness metrics, and 20 fairness-performance trade-off measurements for 8 widely-adopted benchmark tasks. The results of our comprehensive study reveal a series of findings. In particular, we find that (1) the bias mitigation methods significantly decrease the values of all ML performance metrics in a notable proportion of scenarios (42\%$\sim$66\% according to different metrics);
(2) the bias mitigation methods achieve significant fairness improvement in 46\% of scenarios (24\%$\sim$59\% according to different fairness metrics); 
(3) the bias mitigation methods even lead to decreases in both fairness and ML performance in 25\% of scenarios;
(4) the effectiveness of the bias mitigation methods depends on tasks, models, the choice of protected attributes, and the set of metrics used to assess fairness and ML performance; (5) there is no bias mitigation method that can achieve the best trade-off in all the scenarios that we study. The best bias mitigation method that we find outperforms other methods in 30\% of scenarios.

In the future, we plan to develop new bias mitigation methods to tackle the weaknesses of existing methods revealed by our study. Moreover, considering that there is no ``silver bullet’’ bias  mitigation method, we plan to propose a framework that selects the most suitable bias mitigation method for different intended scenarios. In addition, when more methods, evaluation metrics, and datasets are proposed in the future, we plan to replicate this study.

\begin{acks}
Zhenpeng Chen, Federica Sarro, and Mark Harman are supported by the ERC Advanced Grant under the grant number 741278 (EPIC: Evolutionary Program Improvement Collaborators).
Jie M. Zhang is partially supported by the UKRI Trustworthy Autonomous Systems Node in Verifiability, with Grant Award Reference EP/V026801/2.
\end{acks}

\bibliographystyle{ACM-Reference-Format}
\bibliography{tosem-fairness}

%%% -*-BibTeX-*-
%%% Do NOT edit. File created by BibTeX with style
%%% ACM-Reference-Format-Journals [18-Jan-2012].

\begin{thebibliography}{81}

%%% ====================================================================
%%% NOTE TO THE USER: you can override these defaults by providing
%%% customized versions of any of these macros before the \bibliography
%%% command.  Each of them MUST provide its own final punctuation,
%%% except for \shownote{}, \showDOI{}, and \showURL{}.  The latter two
%%% do not use final punctuation, in order to avoid confusing it with
%%% the Web address.
%%%
%%% To suppress output of a particular field, define its macro to expand
%%% to an empty string, or better, \unskip, like this:
%%%
%%% \newcommand{\showDOI}[1]{\unskip}   % LaTeX syntax
%%%
%%% \def \showDOI #1{\unskip}           % plain TeX syntax
%%%
%%% ====================================================================

\ifx \showCODEN    \undefined \def \showCODEN     #1{\unskip}     \fi
\ifx \showDOI      \undefined \def \showDOI       #1{#1}\fi
\ifx \showISBNx    \undefined \def \showISBNx     #1{\unskip}     \fi
\ifx \showISBNxiii \undefined \def \showISBNxiii  #1{\unskip}     \fi
\ifx \showISSN     \undefined \def \showISSN      #1{\unskip}     \fi
\ifx \showLCCN     \undefined \def \showLCCN      #1{\unskip}     \fi
\ifx \shownote     \undefined \def \shownote      #1{#1}          \fi
\ifx \showarticletitle \undefined \def \showarticletitle #1{#1}   \fi
\ifx \showURL      \undefined \def \showURL       {\relax}        \fi
% The following commands are used for tagged output and should be
% invisible to TeX
\providecommand\bibfield[2]{#2}
\providecommand\bibinfo[2]{#2}
\providecommand\natexlab[1]{#1}
\providecommand\showeprint[2][]{arXiv:#2}

\bibitem[\protect\citeauthoryear{??}{adu}{[n.d.]}]%
        {adultdata}
 \bibinfo{year}{[n.d.]}\natexlab{}.
\newblock \bibinfo{title}{The Adult Census Income dataset}.
\newblock
  \bibinfo{howpublished}{\url{https://archive.ics.uci.edu/ml/datasets/adult}}.
\newblock
\newblock
\shownote{Retrieved on September 20, 2021.}


\bibitem[\protect\citeauthoryear{??}{ref}{[n.d.]a}]%
        {refer2}
 \bibinfo{year}{[n.d.]}\natexlab{a}.
\newblock \bibinfo{title}{Applications of scikit-learn}.
\newblock
  \bibinfo{howpublished}{\url{https://numfocus.org/project/scikit-learn\#:~:text=Implementations\%20rely\%20either\%20on\%20vectorized,to\%20analyzing\%20brain\%20imaging\%20data}}.
\newblock
\newblock
\shownote{Retrieved on November 24, 2022.}


\bibitem[\protect\citeauthoryear{??}{ban}{[n.d.]}]%
        {bankdata}
 \bibinfo{year}{[n.d.]}\natexlab{}.
\newblock \bibinfo{title}{The Bank dataset}.
\newblock
  \bibinfo{howpublished}{\url{https://archive.ics.uci.edu/ml/datasets/Bank+Marketing}}.
\newblock
\newblock
\shownote{Retrieved on September 20, 2021.}


\bibitem[\protect\citeauthoryear{??}{com}{[n.d.]}]%
        {compasdata}
 \bibinfo{year}{[n.d.]}\natexlab{}.
\newblock \bibinfo{title}{The Compas dataset}.
\newblock
  \bibinfo{howpublished}{\url{https://github.com/propublica/compas-analysis}}.
\newblock
\newblock
\shownote{Retrieved on September 20, 2021.}


\bibitem[\protect\citeauthoryear{??}{fat}{[n.d.]}]%
        {fategroup}
 \bibinfo{year}{[n.d.]}\natexlab{}.
\newblock \bibinfo{title}{FATE: Fairness, Accountability, Transparency, and
  Ethics in AI}.
\newblock
  \bibinfo{howpublished}{\url{https://www.microsoft.com/en-us/research/theme/fate/}}.
\newblock
\newblock
\shownote{Retrieved on September 20, 2021.}


\bibitem[\protect\citeauthoryear{??}{ger}{[n.d.]}]%
        {germandata}
 \bibinfo{year}{[n.d.]}\natexlab{}.
\newblock \bibinfo{title}{The German Credit dataset}.
\newblock
  \bibinfo{howpublished}{\url{https://archive.ics.uci.edu/ml/datasets/Statlog+\%28German+Credit+Data\%29}}.
\newblock
\newblock
\shownote{Retrieved on September 20, 2021.}


\bibitem[\protect\citeauthoryear{??}{fai}{[n.d.]a}]%
        {fairsmotegit}
 \bibinfo{year}{[n.d.]}\natexlab{a}.
\newblock \bibinfo{title}{The GitHub repository of Fair-SMOTE}.
\newblock
  \bibinfo{howpublished}{\url{https://github.com/joymallyac/Fair-SMOTE/tree/master/Fair-SMOTE}}.
\newblock
\newblock
\shownote{Retrieved on September 20, 2021.}


\bibitem[\protect\citeauthoryear{??}{fai}{[n.d.]b}]%
        {fairwaygit}
 \bibinfo{year}{[n.d.]}\natexlab{b}.
\newblock \bibinfo{title}{The GitHub repository of Fairway}.
\newblock \bibinfo{howpublished}{\url{https://github.com/joymallyac/Fairway}}.
\newblock
\newblock
\shownote{Retrieved on September 20, 2021.}


\bibitem[\protect\citeauthoryear{??}{aif}{[n.d.]}]%
        {aif360}
 \bibinfo{year}{[n.d.]}\natexlab{}.
\newblock \bibinfo{title}{IBM AI Fairness 360}.
\newblock \bibinfo{howpublished}{\url{https://aif360.mybluemix.net}}.
\newblock
\newblock
\shownote{Retrieved on September 20, 2021.}


\bibitem[\protect\citeauthoryear{??}{fai}{[n.d.]c}]%
        {faircase}
 \bibinfo{year}{[n.d.]}\natexlab{c}.
\newblock \bibinfo{title}{Machine Bias}.
\newblock
  \bibinfo{howpublished}{\url{https://www.propublica.org/article/machine-bias-risk-assessments-in-criminal-sentencing}}.
\newblock
\newblock
\shownote{Retrieved on September 20, 2021.}


\bibitem[\protect\citeauthoryear{??}{mep}{[n.d.]}]%
        {mepdata}
 \bibinfo{year}{[n.d.]}\natexlab{}.
\newblock \bibinfo{title}{The Mep dataset}.
\newblock
  \bibinfo{howpublished}{\url{https://meps.ahrq.gov/mepsweb/data_stats/download_data_files_detail.jsp?cboPufNumber=HC-181}}.
\newblock
\newblock
\shownote{Retrieved on September 20, 2021.}


\bibitem[\protect\citeauthoryear{??}{mic}{[n.d.]}]%
        {microsoftai}
 \bibinfo{year}{[n.d.]}\natexlab{}.
\newblock \bibinfo{title}{Microsoft AI principles}.
\newblock
  \bibinfo{howpublished}{\url{https://www.microsoft.com/en-us/ai/responsible-ai?activetab=pivot1\%3aprimaryr6}}.
\newblock
\newblock
\shownote{Retrieved on September 20, 2021.}


\bibitem[\protect\citeauthoryear{??}{ref}{[n.d.]b}]%
        {refer4}
 \bibinfo{year}{[n.d.]}\natexlab{b}.
\newblock \bibinfo{title}{Review into bias in algorithmic decision-making}.
\newblock
  \bibinfo{howpublished}{\url{https://www.gov.uk/government/publications/cdei-publishes-review-into-bias-in-algorithmic-decision-making/main-report-cdei-review-into-bias-in-algorithmic-decision-making}}.
\newblock
\newblock
\shownote{Retrieved on November 24, 2022.}


\bibitem[\protect\citeauthoryear{??}{skl}{[n.d.]}]%
        {sklearn}
 \bibinfo{year}{[n.d.]}\natexlab{}.
\newblock \bibinfo{title}{Scikit-learn}.
\newblock \bibinfo{howpublished}{\url{https://scikit-learn.org}}.
\newblock
\newblock
\shownote{Retrieved on September 20, 2021.}


\bibitem[\protect\citeauthoryear{??}{fai}{[n.d.]d}]%
        {faircase3}
 \bibinfo{year}{[n.d.]}\natexlab{d}.
\newblock \bibinfo{title}{Semantics derived automatically from language corpora
  contain human-like biases}.
\newblock
  \bibinfo{howpublished}{\url{https://www.science.org/doi/10.1126/science.aal4230}}.
\newblock
\newblock
\shownote{Retrieved on September 20, 2021.}


\bibitem[\protect\citeauthoryear{??}{fai}{[n.d.]e}]%
        {faircase2}
 \bibinfo{year}{[n.d.]}\natexlab{e}.
\newblock \bibinfo{title}{Study finds gender and skin-type bias in commercial
  artificial-intelligence systems}.
\newblock
  \bibinfo{howpublished}{\url{https://news.mit.edu/2018/study-finds-gender-skin-type-bias-artificial-intelligence-systems-0212}}.
\newblock
\newblock
\shownote{Retrieved on September 20, 2021.}


\bibitem[\protect\citeauthoryear{??}{fai}{[n.d.]f}]%
        {faircase1}
 \bibinfo{year}{[n.d.]}\natexlab{f}.
\newblock \bibinfo{title}{When good algorithms go sexist: why and how to
  advance AI gender equity}.
\newblock
  \bibinfo{howpublished}{\url{https://ssir.org/articles/entry/when_good_algorithms_go_sexist_why_and_how_to_advance_ai_gender_equity}}.
\newblock
\newblock
\shownote{Retrieved on September 20, 2021.}


\bibitem[\protect\citeauthoryear{??}{ref}{[n.d.]c}]%
        {refer1}
 \bibinfo{year}{[n.d.]}\natexlab{c}.
\newblock \bibinfo{title}{Who is using scikit-learn?}
\newblock
  \bibinfo{howpublished}{\url{https://scikit-learn.org/stable/testimonials/testimonials.html\#id8}}.
\newblock
\newblock
\shownote{Retrieved on November 24, 2022.}


\bibitem[\protect\citeauthoryear{Aggarwal, Lohia, Nagar, Dey, and
  Saha}{Aggarwal et~al\mbox{.}}{2019}]%
        {sigsoftAggarwalLNDS19}
\bibfield{author}{\bibinfo{person}{Aniya Aggarwal}, \bibinfo{person}{Pranay
  Lohia}, \bibinfo{person}{Seema Nagar}, \bibinfo{person}{Kuntal Dey}, {and}
  \bibinfo{person}{Diptikalyan Saha}.} \bibinfo{year}{2019}\natexlab{}.
\newblock \showarticletitle{Black box fairness testing of machine learning
  models}. In \bibinfo{booktitle}{\emph{Proceedings of the {ACM} Joint Meeting
  on European Software Engineering Conference and Symposium on the Foundations
  of Software Engineering, {ESEC/FSE} 2019}}. \bibinfo{pages}{625--635}.
\newblock


\bibitem[\protect\citeauthoryear{Alabi, Abdulsalam, Ogundokun, and
  Arowolo}{Alabi et~al\mbox{.}}{2020}]%
        {alabi2020credit}
\bibfield{author}{\bibinfo{person}{Kayode~Omotosho Alabi},
  \bibinfo{person}{Sulaiman~Olaniyi Abdulsalam},
  \bibinfo{person}{Roseline~Oluwaseun Ogundokun}, {and}
  \bibinfo{person}{Micheal~Olaolu Arowolo}.} \bibinfo{year}{2020}\natexlab{}.
\newblock \showarticletitle{Credit risk prediction in commercial bank using
  chi-square with SVM-RBF}. In \bibinfo{booktitle}{\emph{International
  Conference on Information and Communication Technology and Applications}}.
  \bibinfo{pages}{158--169}.
\newblock


\bibitem[\protect\citeauthoryear{Arcuri and Briand}{Arcuri and Briand}{2011}]%
        {icseArcuriB11}
\bibfield{author}{\bibinfo{person}{Andrea Arcuri} {and}
  \bibinfo{person}{Lionel~C. Briand}.} \bibinfo{year}{2011}\natexlab{}.
\newblock \showarticletitle{A practical guide for using statistical tests to
  assess randomized algorithms in software engineering}. In
  \bibinfo{booktitle}{\emph{Proceedings of the 33rd International Conference on
  Software Engineering, {ICSE} 2011}}. \bibinfo{pages}{1--10}.
\newblock


\bibitem[\protect\citeauthoryear{Berk, Heidari, Jabbari, Kearns, and Roth}{Berk
  et~al\mbox{.}}{2021}]%
        {berk2021fairness}
\bibfield{author}{\bibinfo{person}{Richard Berk}, \bibinfo{person}{Hoda
  Heidari}, \bibinfo{person}{Shahin Jabbari}, \bibinfo{person}{Michael Kearns},
  {and} \bibinfo{person}{Aaron Roth}.} \bibinfo{year}{2021}\natexlab{}.
\newblock \showarticletitle{Fairness in criminal justice risk assessments: The
  state of the art}.
\newblock \bibinfo{journal}{\emph{Sociological Methods \& Research}}
  \bibinfo{volume}{50}, \bibinfo{number}{1} (\bibinfo{year}{2021}),
  \bibinfo{pages}{3--44}.
\newblock


\bibitem[\protect\citeauthoryear{Biswas and Rajan}{Biswas and Rajan}{2020}]%
        {biswas2020machine}
\bibfield{author}{\bibinfo{person}{Sumon Biswas} {and} \bibinfo{person}{Hridesh
  Rajan}.} \bibinfo{year}{2020}\natexlab{}.
\newblock \showarticletitle{Do the machine learning models on a crowd sourced
  platform exhibit bias? An empirical study on model fairness}. In
  \bibinfo{booktitle}{\emph{Proceedings of the 28th ACM Joint Meeting on
  European Software Engineering Conference and Symposium on the Foundations of
  Software Engineering, ESEC/FSE 2020}}. \bibinfo{pages}{642--653}.
\newblock


\bibitem[\protect\citeauthoryear{Biswas and Rajan}{Biswas and Rajan}{2021}]%
        {sigsoftBiswasR21}
\bibfield{author}{\bibinfo{person}{Sumon Biswas} {and} \bibinfo{person}{Hridesh
  Rajan}.} \bibinfo{year}{2021}\natexlab{}.
\newblock \showarticletitle{Fair preprocessing: towards understanding
  compositional fairness of data transformers in machine learning pipeline}. In
  \bibinfo{booktitle}{\emph{Proceedings of the 29th {ACM} Joint European
  Software Engineering Conference and Symposium on the Foundations of Software
  Engineering, {ESEC/FSE} 2021}}. \bibinfo{pages}{981--993}.
\newblock


\bibitem[\protect\citeauthoryear{Briand, Emam, and Morasca}{Briand
  et~al\mbox{.}}{1996}]%
        {eseBriandEM96}
\bibfield{author}{\bibinfo{person}{Lionel~C. Briand},
  \bibinfo{person}{Khaled~El Emam}, {and} \bibinfo{person}{Sandro Morasca}.}
  \bibinfo{year}{1996}\natexlab{}.
\newblock \showarticletitle{On the application of measurement theory in
  software engineering}.
\newblock \bibinfo{journal}{\emph{Empirical Software Engineering}}
  \bibinfo{volume}{1}, \bibinfo{number}{1} (\bibinfo{year}{1996}),
  \bibinfo{pages}{61--88}.
\newblock


\bibitem[\protect\citeauthoryear{Brun and Meliou}{Brun and Meliou}{2018}]%
        {sigsoftBrunM18}
\bibfield{author}{\bibinfo{person}{Yuriy Brun} {and} \bibinfo{person}{Alexandra
  Meliou}.} \bibinfo{year}{2018}\natexlab{}.
\newblock \showarticletitle{Software fairness}. In
  \bibinfo{booktitle}{\emph{Proceedings of the 2018 {ACM} Joint Meeting on
  European Software Engineering Conference and Symposium on the Foundations of
  Software Engineering, {ESEC/FSE} 2018}}. \bibinfo{pages}{754--759}.
\newblock


\bibitem[\protect\citeauthoryear{Calders, Kamiran, and Pechenizkiy}{Calders
  et~al\mbox{.}}{2009}]%
        {icdmCaldersKP09}
\bibfield{author}{\bibinfo{person}{Toon Calders}, \bibinfo{person}{Faisal
  Kamiran}, {and} \bibinfo{person}{Mykola Pechenizkiy}.}
  \bibinfo{year}{2009}\natexlab{}.
\newblock \showarticletitle{Building classifiers with independency
  constraints}. In \bibinfo{booktitle}{\emph{Proceedings of the 2009, {IEEE}
  International Conference on Data Mining}}. \bibinfo{pages}{13--18}.
\newblock


\bibitem[\protect\citeauthoryear{Calders and Verwer}{Calders and
  Verwer}{2010}]%
        {datamineCaldersV10}
\bibfield{author}{\bibinfo{person}{Toon Calders} {and} \bibinfo{person}{Sicco
  Verwer}.} \bibinfo{year}{2010}\natexlab{}.
\newblock \showarticletitle{Three naive Bayes approaches for
  discrimination-free classification}.
\newblock \bibinfo{journal}{\emph{Data Mining and Knowledge Discovery}}
  \bibinfo{volume}{21}, \bibinfo{number}{2} (\bibinfo{year}{2010}),
  \bibinfo{pages}{277--292}.
\newblock


\bibitem[\protect\citeauthoryear{Celis, Huang, Keswani, and Vishnoi}{Celis
  et~al\mbox{.}}{2019}]%
        {mfcpaper}
\bibfield{author}{\bibinfo{person}{L.~Elisa Celis}, \bibinfo{person}{Lingxiao
  Huang}, \bibinfo{person}{Vijay Keswani}, {and} \bibinfo{person}{Nisheeth~K.
  Vishnoi}.} \bibinfo{year}{2019}\natexlab{}.
\newblock \showarticletitle{Classification with fairness constraints: a
  meta-algorithm wit provable guarantees}. In
  \bibinfo{booktitle}{\emph{Proceedings of the Conference on Fairness,
  Accountability, and Transparency, FAT* 2019}}. \bibinfo{pages}{319--328}.
\newblock


\bibitem[\protect\citeauthoryear{Chakraborty, Majumder, and
  Menzies}{Chakraborty et~al\mbox{.}}{2021}]%
        {fairsmotepaper}
\bibfield{author}{\bibinfo{person}{Joymallya Chakraborty},
  \bibinfo{person}{Suvodeep Majumder}, {and} \bibinfo{person}{Tim Menzies}.}
  \bibinfo{year}{2021}\natexlab{}.
\newblock \showarticletitle{Bias in machine learning software: why? how? what
  to do?}. In \bibinfo{booktitle}{\emph{Proceedings of the 29th {ACM} Joint
  European Software Engineering Conference and Symposium on the Foundations of
  Software Engineering, {ESEC/FSE} 2021}}. \bibinfo{pages}{429--440}.
\newblock


\bibitem[\protect\citeauthoryear{Chakraborty, Majumder, Yu, and
  Menzies}{Chakraborty et~al\mbox{.}}{2020a}]%
        {fairwaypaper}
\bibfield{author}{\bibinfo{person}{Joymallya Chakraborty},
  \bibinfo{person}{Suvodeep Majumder}, \bibinfo{person}{Zhe Yu}, {and}
  \bibinfo{person}{Tim Menzies}.} \bibinfo{year}{2020}\natexlab{a}.
\newblock \showarticletitle{Fairway: a way to build fair {ML} software}. In
  \bibinfo{booktitle}{\emph{Proceedings of the 28th {ACM} Joint European
  Software Engineering Conference and Symposium on the Foundations of Software
  Engineering, {ESEC/FSE} 2020}}. \bibinfo{pages}{654--665}.
\newblock


\bibitem[\protect\citeauthoryear{Chakraborty, Peng, and Menzies}{Chakraborty
  et~al\mbox{.}}{2020b}]%
        {kbseChakrabortyPM20}
\bibfield{author}{\bibinfo{person}{Joymallya Chakraborty},
  \bibinfo{person}{Kewen Peng}, {and} \bibinfo{person}{Tim Menzies}.}
  \bibinfo{year}{2020}\natexlab{b}.
\newblock \showarticletitle{Making fair {ML} software using trustworthy
  explanation}. In \bibinfo{booktitle}{\emph{Proceedings of the 35th {IEEE/ACM}
  International Conference on Automated Software Engineering, {ASE} 2020}}.
  \bibinfo{pages}{1229--1233}.
\newblock


\bibitem[\protect\citeauthoryear{Chen, Cao, Lu, Mei, and Liu}{Chen
  et~al\mbox{.}}{2019}]%
        {sigsoftChenCLML19}
\bibfield{author}{\bibinfo{person}{Zhenpeng Chen}, \bibinfo{person}{Yanbin
  Cao}, \bibinfo{person}{Xuan Lu}, \bibinfo{person}{Qiaozhu Mei}, {and}
  \bibinfo{person}{Xuanzhe Liu}.} \bibinfo{year}{2019}\natexlab{}.
\newblock \showarticletitle{SEntiMoji: an emoji-powered learning approach for
  sentiment analysis in software engineering}. In
  \bibinfo{booktitle}{\emph{Proceedings of the {ACM} Joint Meeting on European
  Software Engineering Conference and Symposium on the Foundations of Software
  Engineering, {ESECFSE} 2019}}. \bibinfo{pages}{841--852}.
\newblock


\bibitem[\protect\citeauthoryear{Chen, Cao, Yao, Lu, Peng, Mei, and Liu}{Chen
  et~al\mbox{.}}{2021}]%
        {tosemChenCYLPML21}
\bibfield{author}{\bibinfo{person}{Zhenpeng Chen}, \bibinfo{person}{Yanbin
  Cao}, \bibinfo{person}{Huihan Yao}, \bibinfo{person}{Xuan Lu},
  \bibinfo{person}{Xin Peng}, \bibinfo{person}{Hong Mei}, {and}
  \bibinfo{person}{Xuanzhe Liu}.} \bibinfo{year}{2021}\natexlab{}.
\newblock \showarticletitle{Emoji-powered sentiment and emotion detection from
  software developers' communication data}.
\newblock \bibinfo{journal}{\emph{ACM Transactions on Software Engineering and
  Methodology}} \bibinfo{volume}{30}, \bibinfo{number}{2}
  (\bibinfo{year}{2021}), \bibinfo{pages}{18:1--18:48}.
\newblock


\bibitem[\protect\citeauthoryear{Chen, Zhang, Hort, Sarro, and Harman}{Chen
  et~al\mbox{.}}{2022a}]%
        {Dabs220710223}
\bibfield{author}{\bibinfo{person}{Zhenpeng Chen}, \bibinfo{person}{Jie~M.
  Zhang}, \bibinfo{person}{Max Hort}, \bibinfo{person}{Federica Sarro}, {and}
  \bibinfo{person}{Mark Harman}.} \bibinfo{year}{2022}\natexlab{a}.
\newblock \showarticletitle{Fairness testing: A comprehensive survey and
  analysis of trends}.
\newblock \bibinfo{journal}{\emph{CoRR}}  \bibinfo{volume}{abs/2207.10223}
  (\bibinfo{year}{2022}).
\newblock


\bibitem[\protect\citeauthoryear{Chen, Zhang, Sarro, and Harman}{Chen
  et~al\mbox{.}}{2022b}]%
        {sigsoftChenZSH22}
\bibfield{author}{\bibinfo{person}{Zhenpeng Chen}, \bibinfo{person}{Jie~M.
  Zhang}, \bibinfo{person}{Federica Sarro}, {and} \bibinfo{person}{Mark
  Harman}.} \bibinfo{year}{2022}\natexlab{b}.
\newblock \showarticletitle{{MAAT:} A novel ensemble approach to addressing
  fairness and performance bugs for machine learning software}. In
  \bibinfo{booktitle}{\emph{Proceedings of the 30th {ACM} Joint European
  Software Engineering Conference and Symposium on the Foundations of Software
  Engineering, {ESEC/FSE} 2022}}. \bibinfo{pages}{1122--1134}.
\newblock


\bibitem[\protect\citeauthoryear{Chen, Zhang, Sarro, and Harman}{Chen
  et~al\mbox{.}}{2023}]%
        {githublink}
\bibfield{author}{\bibinfo{person}{Zhenpeng Chen}, \bibinfo{person}{Jie~M.
  Zhang}, \bibinfo{person}{Federica Sarro}, {and} \bibinfo{person}{Mark
  Harman}.} \bibinfo{year}{2023}\natexlab{}.
\newblock \bibinfo{title}{Replication package}.
\newblock
  \bibinfo{howpublished}{\url{https://github.com/chenzhenpeng18/TOSEM23-BiasMitigationStudy}}.
\newblock
\newblock
\shownote{Retrieved on January 14, 2023.}


\bibitem[\protect\citeauthoryear{Chicco and Jurman}{Chicco and Jurman}{2020}]%
        {chicco2020advantages}
\bibfield{author}{\bibinfo{person}{Davide Chicco} {and}
  \bibinfo{person}{Giuseppe Jurman}.} \bibinfo{year}{2020}\natexlab{}.
\newblock \showarticletitle{The advantages of the Matthews correlation
  coefficient (MCC) over F1 score and accuracy in binary classification
  evaluation}.
\newblock \bibinfo{journal}{\emph{BMC genomics}} \bibinfo{volume}{21},
  \bibinfo{number}{1} (\bibinfo{year}{2020}), \bibinfo{pages}{1--13}.
\newblock


\bibitem[\protect\citeauthoryear{Chouldechova}{Chouldechova}{2017}]%
        {chouldechova2017fair}
\bibfield{author}{\bibinfo{person}{Alexandra Chouldechova}.}
  \bibinfo{year}{2017}\natexlab{}.
\newblock \showarticletitle{Fair prediction with disparate impact: a study of
  bias in recidivism prediction instruments}.
\newblock \bibinfo{journal}{\emph{Big data}} \bibinfo{volume}{5},
  \bibinfo{number}{2} (\bibinfo{year}{2017}), \bibinfo{pages}{153--163}.
\newblock


\bibitem[\protect\citeauthoryear{Cohen}{Cohen}{2013}]%
        {cohen2013statistical}
\bibfield{author}{\bibinfo{person}{Jacob Cohen}.}
  \bibinfo{year}{2013}\natexlab{}.
\newblock \bibinfo{booktitle}{\emph{Statistical power analysis for the
  behavioral sciences}}.
\newblock \bibinfo{publisher}{Academic press}.
\newblock


\bibitem[\protect\citeauthoryear{du~Pin~Calmon, Wei, Vinzamuri, Ramamurthy, and
  Varshney}{du~Pin~Calmon et~al\mbox{.}}{2017}]%
        {oppaper}
\bibfield{author}{\bibinfo{person}{Fl{\'{a}}vio du Pin~Calmon},
  \bibinfo{person}{Dennis Wei}, \bibinfo{person}{Bhanukiran Vinzamuri},
  \bibinfo{person}{Karthikeyan~Natesan Ramamurthy}, {and}
  \bibinfo{person}{Kush~R. Varshney}.} \bibinfo{year}{2017}\natexlab{}.
\newblock \showarticletitle{Optimized pre-processing for discrimination
  prevention}. In \bibinfo{booktitle}{\emph{Proceedings of the Annual
  Conference on Neural Information Processing Systems 2017, NIPS 2017}}.
  \bibinfo{pages}{3992--4001}.
\newblock


\bibitem[\protect\citeauthoryear{Feldman, Friedler, Moeller, Scheidegger, and
  Venkatasubramanian}{Feldman et~al\mbox{.}}{2015a}]%
        {kddFeldmanFMSV15}
\bibfield{author}{\bibinfo{person}{Michael Feldman},
  \bibinfo{person}{Sorelle~A. Friedler}, \bibinfo{person}{John Moeller},
  \bibinfo{person}{Carlos Scheidegger}, {and} \bibinfo{person}{Suresh
  Venkatasubramanian}.} \bibinfo{year}{2015}\natexlab{a}.
\newblock \showarticletitle{Certifying and removing disparate impact}. In
  \bibinfo{booktitle}{\emph{Proceedings of the 21th {ACM} {SIGKDD}
  International Conference on Knowledge Discovery and Data Mining}}.
  \bibinfo{pages}{259--268}.
\newblock


\bibitem[\protect\citeauthoryear{Feldman, Friedler, Moeller, Scheidegger, and
  Venkatasubramanian}{Feldman et~al\mbox{.}}{2015b}]%
        {DIpaper}
\bibfield{author}{\bibinfo{person}{Michael Feldman},
  \bibinfo{person}{Sorelle~A. Friedler}, \bibinfo{person}{John Moeller},
  \bibinfo{person}{Carlos Scheidegger}, {and} \bibinfo{person}{Suresh
  Venkatasubramanian}.} \bibinfo{year}{2015}\natexlab{b}.
\newblock \showarticletitle{Certifying and removing disparate impact}. In
  \bibinfo{booktitle}{\emph{Proceedings of the 21th {ACM} {SIGKDD}
  International Conference on Knowledge Discovery and Data Mining, KDD 2015}}.
  \bibinfo{pages}{259--268}.
\newblock


\bibitem[\protect\citeauthoryear{Finkelstein, Harman, Mansouri, Ren, and
  Zhang}{Finkelstein et~al\mbox{.}}{2008}]%
        {afetal:re08}
\bibfield{author}{\bibinfo{person}{Anthony Finkelstein}, \bibinfo{person}{Mark
  Harman}, \bibinfo{person}{Afshin Mansouri}, \bibinfo{person}{Jian Ren}, {and}
  \bibinfo{person}{Yuanyuan Zhang}.} \bibinfo{year}{2008}\natexlab{}.
\newblock \showarticletitle{Fairness analysis in requirements assignments}. In
  \bibinfo{booktitle}{\emph{Proceedings of the 16th {IEEE} International
  Requirements Engineering Conference}}. \bibinfo{pages}{115--124}.
\newblock


\bibitem[\protect\citeauthoryear{Galhotra, Brun, and Meliou}{Galhotra
  et~al\mbox{.}}{2017}]%
        {sigsoftGalhotraBM17}
\bibfield{author}{\bibinfo{person}{Sainyam Galhotra}, \bibinfo{person}{Yuriy
  Brun}, {and} \bibinfo{person}{Alexandra Meliou}.}
  \bibinfo{year}{2017}\natexlab{}.
\newblock \showarticletitle{Fairness testing: testing software for
  discrimination}. In \bibinfo{booktitle}{\emph{Proceedings of the 2017 11th
  Joint Meeting on Foundations of Software Engineering, {ESEC/FSE} 2017}}.
  \bibinfo{pages}{498--510}.
\newblock


\bibitem[\protect\citeauthoryear{Ghasemi, Anvari, Atapour, Stephen~Wormith,
  Stockdale, and Spiteri}{Ghasemi et~al\mbox{.}}{2021}]%
        {ghasemi2021application}
\bibfield{author}{\bibinfo{person}{Mehdi Ghasemi}, \bibinfo{person}{Daniel
  Anvari}, \bibinfo{person}{Mahshid Atapour}, \bibinfo{person}{J
  Stephen~Wormith}, \bibinfo{person}{Keira~C Stockdale}, {and}
  \bibinfo{person}{Raymond~J Spiteri}.} \bibinfo{year}{2021}\natexlab{}.
\newblock \showarticletitle{The Application of Machine Learning to a General
  Risk--Need Assessment Instrument in the Prediction of Criminal Recidivism}.
\newblock \bibinfo{journal}{\emph{Criminal Justice and Behavior}}
  \bibinfo{volume}{48}, \bibinfo{number}{4} (\bibinfo{year}{2021}),
  \bibinfo{pages}{518--538}.
\newblock


\bibitem[\protect\citeauthoryear{Hardt, Price, and Srebro}{Hardt
  et~al\mbox{.}}{2016}]%
        {EOpaper}
\bibfield{author}{\bibinfo{person}{Moritz Hardt}, \bibinfo{person}{Eric Price},
  {and} \bibinfo{person}{Nati Srebro}.} \bibinfo{year}{2016}\natexlab{}.
\newblock \showarticletitle{Equality of opportunity in supervised learning}. In
  \bibinfo{booktitle}{\emph{Proceedings of the Annual Conference on Neural
  Information Processing Systems 2016, NIPS 2016}}.
  \bibinfo{pages}{3315--3323}.
\newblock


\bibitem[\protect\citeauthoryear{Hort, Chen, Zhang, Sarro, and Harman}{Hort
  et~al\mbox{.}}{2022}]%
        {DBcorrabs220707068}
\bibfield{author}{\bibinfo{person}{Max Hort}, \bibinfo{person}{Zhenpeng Chen},
  \bibinfo{person}{Jie~M. Zhang}, \bibinfo{person}{Federica Sarro}, {and}
  \bibinfo{person}{Mark Harman}.} \bibinfo{year}{2022}\natexlab{}.
\newblock \showarticletitle{Bias mitigation for machine learning classifiers:
  {A} comprehensive survey}.
\newblock \bibinfo{journal}{\emph{CoRR}}  \bibinfo{volume}{abs/2207.07068}
  (\bibinfo{year}{2022}).
\newblock


\bibitem[\protect\citeauthoryear{Hort and Sarro}{Hort and Sarro}{2021}]%
        {maxasefairness}
\bibfield{author}{\bibinfo{person}{Max Hort} {and} \bibinfo{person}{Federica
  Sarro}.} \bibinfo{year}{2021}\natexlab{}.
\newblock \showarticletitle{Did you do your homework? Raising awareness on
  software fairness and discrimination}. In
  \bibinfo{booktitle}{\emph{Proceedings of the 36th IEEE/ACM International
  Conference on Automated Software Engineering, ASE 2021}}.
\newblock


\bibitem[\protect\citeauthoryear{Hort, Zhang, Sarro, and Harman}{Hort
  et~al\mbox{.}}{2021}]%
        {sigsoftHortZSH21}
\bibfield{author}{\bibinfo{person}{Max Hort}, \bibinfo{person}{Jie~M. Zhang},
  \bibinfo{person}{Federica Sarro}, {and} \bibinfo{person}{Mark Harman}.}
  \bibinfo{year}{2021}\natexlab{}.
\newblock \showarticletitle{Fairea: a model behaviour mutation approach to
  benchmarking bias mitigation methods}. In
  \bibinfo{booktitle}{\emph{Proceedings of the 29th {ACM} Joint European
  Software Engineering Conference and Symposium on the Foundations of Software
  Engineering, Athens, ESEC/FSE 2021}}. \bibinfo{pages}{994--1006}.
\newblock


\bibitem[\protect\citeauthoryear{Kamiran and Calders}{Kamiran and
  Calders}{2011}]%
        {rewpaper}
\bibfield{author}{\bibinfo{person}{Faisal Kamiran} {and} \bibinfo{person}{Toon
  Calders}.} \bibinfo{year}{2011}\natexlab{}.
\newblock \showarticletitle{Data preprocessing techniques for classification
  without discrimination}.
\newblock \bibinfo{journal}{\emph{Knowledge and Information Systems}}
  \bibinfo{volume}{33}, \bibinfo{number}{1} (\bibinfo{year}{2011}),
  \bibinfo{pages}{1--33}.
\newblock


\bibitem[\protect\citeauthoryear{Kamiran, Calders, and Pechenizkiy}{Kamiran
  et~al\mbox{.}}{2010}]%
        {icdmKamiranCP10}
\bibfield{author}{\bibinfo{person}{Faisal Kamiran}, \bibinfo{person}{Toon
  Calders}, {and} \bibinfo{person}{Mykola Pechenizkiy}.}
  \bibinfo{year}{2010}\natexlab{}.
\newblock \showarticletitle{Discrimination Aware Decision Tree Learning}. In
  \bibinfo{booktitle}{\emph{Proceedings of the 10th {IEEE} International
  Conference on Data Mining, ICDM 2010}}. \bibinfo{pages}{869--874}.
\newblock


\bibitem[\protect\citeauthoryear{Kamiran, Karim, and Zhang}{Kamiran
  et~al\mbox{.}}{2012}]%
        {ROCpaper}
\bibfield{author}{\bibinfo{person}{Faisal Kamiran}, \bibinfo{person}{Asim
  Karim}, {and} \bibinfo{person}{Xiangliang Zhang}.}
  \bibinfo{year}{2012}\natexlab{}.
\newblock \showarticletitle{Decision theory for discrimination-aware
  classification}. In \bibinfo{booktitle}{\emph{Proceedings of the 12th {IEEE}
  International Conference on Data Mining, {ICDM} 2012}}.
  \bibinfo{pages}{924--929}.
\newblock


\bibitem[\protect\citeauthoryear{Kamiran, Mansha, Karim, and Zhang}{Kamiran
  et~al\mbox{.}}{2018}]%
        {isciKamiranMKZ18}
\bibfield{author}{\bibinfo{person}{Faisal Kamiran}, \bibinfo{person}{Sameen
  Mansha}, \bibinfo{person}{Asim Karim}, {and} \bibinfo{person}{Xiangliang
  Zhang}.} \bibinfo{year}{2018}\natexlab{}.
\newblock \showarticletitle{Exploiting reject option in classification for
  social discrimination control}.
\newblock \bibinfo{journal}{\emph{Information Science}}  \bibinfo{volume}{425}
  (\bibinfo{year}{2018}), \bibinfo{pages}{18--33}.
\newblock


\bibitem[\protect\citeauthoryear{Kamishima, Akaho, Asoh, and Sakuma}{Kamishima
  et~al\mbox{.}}{2012}]%
        {PRpaper}
\bibfield{author}{\bibinfo{person}{Toshihiro Kamishima},
  \bibinfo{person}{Shotaro Akaho}, \bibinfo{person}{Hideki Asoh}, {and}
  \bibinfo{person}{Jun Sakuma}.} \bibinfo{year}{2012}\natexlab{}.
\newblock \showarticletitle{Fairness-aware classifier with prejudice remover
  regularizer}. In \bibinfo{booktitle}{\emph{Proceedings of the European
  Conference on Machine Learning and Knowledge Discovery in Databases,
  {ECML/PKDD} 2012}}. \bibinfo{pages}{35--50}.
\newblock


\bibitem[\protect\citeauthoryear{Mangani}{Mangani}{2004}]%
        {mangani2004online}
\bibfield{author}{\bibinfo{person}{Andrea Mangani}.}
  \bibinfo{year}{2004}\natexlab{}.
\newblock \showarticletitle{Online advertising: Pay-per-view versus
  pay-per-click}.
\newblock \bibinfo{journal}{\emph{Journal of Revenue and Pricing Management}}
  \bibinfo{volume}{2}, \bibinfo{number}{4} (\bibinfo{year}{2004}),
  \bibinfo{pages}{295--302}.
\newblock


\bibitem[\protect\citeauthoryear{Mann and Whitney}{Mann and Whitney}{1947}]%
        {mann1947test}
\bibfield{author}{\bibinfo{person}{Henry~B Mann} {and}
  \bibinfo{person}{Donald~R Whitney}.} \bibinfo{year}{1947}\natexlab{}.
\newblock \showarticletitle{On a test of whether one of two random variables is
  stochastically larger than the other}.
\newblock \bibinfo{journal}{\emph{The Annals of Mathematical Statistics}}
  (\bibinfo{year}{1947}), \bibinfo{pages}{50--60}.
\newblock


\bibitem[\protect\citeauthoryear{Mehrabi, Morstatter, Saxena, Lerman, and
  Galstyan}{Mehrabi et~al\mbox{.}}{2021}]%
        {csurMehrabiMSLG21}
\bibfield{author}{\bibinfo{person}{Ninareh Mehrabi}, \bibinfo{person}{Fred
  Morstatter}, \bibinfo{person}{Nripsuta Saxena}, \bibinfo{person}{Kristina
  Lerman}, {and} \bibinfo{person}{Aram Galstyan}.}
  \bibinfo{year}{2021}\natexlab{}.
\newblock \showarticletitle{A survey on bias and fairness in machine learning}.
\newblock \bibinfo{journal}{\emph{Comput. Surveys}} \bibinfo{volume}{54},
  \bibinfo{number}{6} (\bibinfo{year}{2021}), \bibinfo{pages}{115:1--115:35}.
\newblock


\bibitem[\protect\citeauthoryear{Moussa, Guizzo, and Sarro}{Moussa
  et~al\mbox{.}}{2022}]%
        {esemMoussaGS22}
\bibfield{author}{\bibinfo{person}{Rebecca Moussa}, \bibinfo{person}{Giovani
  Guizzo}, {and} \bibinfo{person}{Federica Sarro}.}
  \bibinfo{year}{2022}\natexlab{}.
\newblock \showarticletitle{{MEG:} Multi-objective ensemble generation for
  software defect prediction}. In \bibinfo{booktitle}{\emph{Proceedings of the
  {ACM} / {IEEE} International Symposium on Empirical Software Engineering and
  Measurement, {ESEM} 2022}}. \bibinfo{pages}{159--170}.
\newblock


\bibitem[\protect\citeauthoryear{Myers, Well, and Lorch~Jr}{Myers
  et~al\mbox{.}}{2013}]%
        {myers2013research}
\bibfield{author}{\bibinfo{person}{Jerome~L Myers}, \bibinfo{person}{Arnold~D
  Well}, {and} \bibinfo{person}{Robert~F Lorch~Jr}.}
  \bibinfo{year}{2013}\natexlab{}.
\newblock \bibinfo{booktitle}{\emph{Research design and statistical analysis}}.
\newblock \bibinfo{publisher}{Routledge}.
\newblock


\bibitem[\protect\citeauthoryear{Novielli, Calefato, Dongiovanni, Girardi, and
  Lanubile}{Novielli et~al\mbox{.}}{2020}]%
        {msrNovielliCDGL20}
\bibfield{author}{\bibinfo{person}{Nicole Novielli}, \bibinfo{person}{Fabio
  Calefato}, \bibinfo{person}{Davide Dongiovanni}, \bibinfo{person}{Daniela
  Girardi}, {and} \bibinfo{person}{Filippo Lanubile}.}
  \bibinfo{year}{2020}\natexlab{}.
\newblock \showarticletitle{Can we use SE-specific sentiment analysis tools in
  a cross-platform setting?}. In \bibinfo{booktitle}{\emph{Proceedings of the
  17th International Conference on Mining Software Repositories, MSR 2020}}.
  \bibinfo{pages}{158--168}.
\newblock


\bibitem[\protect\citeauthoryear{Novielli, Girardi, and Lanubile}{Novielli
  et~al\mbox{.}}{2018}]%
        {NovielliGL08}
\bibfield{author}{\bibinfo{person}{Nicole Novielli}, \bibinfo{person}{Daniela
  Girardi}, {and} \bibinfo{person}{Filippo Lanubile}.}
  \bibinfo{year}{2018}\natexlab{}.
\newblock \showarticletitle{A benchmark study on sentiment analysis for
  software engineering research}. In \bibinfo{booktitle}{\emph{Proceedings of
  the 15th International Conference on Mining Software Repositories, {MSR}
  2018}}. \bibinfo{pages}{364--375}.
\newblock


\bibitem[\protect\citeauthoryear{Pessach and Shmueli}{Pessach and
  Shmueli}{2022}]%
        {cscfairness22}
\bibfield{author}{\bibinfo{person}{D. Pessach} {and} \bibinfo{person}{E.
  Shmueli}.} \bibinfo{year}{2022}\natexlab{}.
\newblock \showarticletitle{A review on fairness in machine learning}.
\newblock \bibinfo{journal}{\emph{Comput. Surveys}} \bibinfo{volume}{55},
  \bibinfo{number}{3} (\bibinfo{year}{2022}).
\newblock


\bibitem[\protect\citeauthoryear{Pleiss, Raghavan, Wu, Kleinberg, and
  Weinberger}{Pleiss et~al\mbox{.}}{2017}]%
        {COpaper}
\bibfield{author}{\bibinfo{person}{Geoff Pleiss}, \bibinfo{person}{Manish
  Raghavan}, \bibinfo{person}{Felix Wu}, \bibinfo{person}{Jon~M. Kleinberg},
  {and} \bibinfo{person}{Kilian~Q. Weinberger}.}
  \bibinfo{year}{2017}\natexlab{}.
\newblock \showarticletitle{On fairness and calibration}. In
  \bibinfo{booktitle}{\emph{Proceedings of the Annual Conference on Neural
  Information Processing Systems 2017, NIPS 2017}}.
  \bibinfo{pages}{5680--5689}.
\newblock


\bibitem[\protect\citeauthoryear{Rodr{\'{\i}}guez, Herraiz, Harrison, Dolado,
  and Riquelme}{Rodr{\'{\i}}guez et~al\mbox{.}}{2014}]%
        {RodriguezHHDR14}
\bibfield{author}{\bibinfo{person}{Daniel Rodr{\'{\i}}guez},
  \bibinfo{person}{Israel Herraiz}, \bibinfo{person}{Rachel Harrison},
  \bibinfo{person}{Jos{\'{e}}~Javier Dolado}, {and}
  \bibinfo{person}{Jos{\'{e}}~C. Riquelme}.} \bibinfo{year}{2014}\natexlab{}.
\newblock \showarticletitle{Preliminary comparison of techniques for dealing
  with imbalance in software defect prediction}. In
  \bibinfo{booktitle}{\emph{Proceedings of the 18th International Conference on
  Evaluation and Assessment in Software Engineering, {EASE} 2014}}.
  \bibinfo{pages}{43:1--43:10}.
\newblock


\bibitem[\protect\citeauthoryear{Sarker}{Sarker}{2021}]%
        {sarker2021machine}
\bibfield{author}{\bibinfo{person}{Iqbal~H Sarker}.}
  \bibinfo{year}{2021}\natexlab{}.
\newblock \showarticletitle{Machine learning: Algorithms, real-world
  applications and research directions}.
\newblock \bibinfo{journal}{\emph{SN Computer Science}} \bibinfo{volume}{2},
  \bibinfo{number}{3} (\bibinfo{year}{2021}), \bibinfo{pages}{1--21}.
\newblock


\bibitem[\protect\citeauthoryear{Sawilowsky}{Sawilowsky}{2009}]%
        {sawilowsky2009new}
\bibfield{author}{\bibinfo{person}{Shlomo~S Sawilowsky}.}
  \bibinfo{year}{2009}\natexlab{}.
\newblock \showarticletitle{New effect size rules of thumb}.
\newblock \bibinfo{journal}{\emph{Journal of Modern Applied Statistical
  Methods}} \bibinfo{volume}{8}, \bibinfo{number}{2} (\bibinfo{year}{2009}),
  \bibinfo{pages}{26}.
\newblock


\bibitem[\protect\citeauthoryear{Sebastiani}{Sebastiani}{2002}]%
        {SebastianiMachine}
\bibfield{author}{\bibinfo{person}{Fabrizio Sebastiani}.}
  \bibinfo{year}{2002}\natexlab{}.
\newblock \showarticletitle{Machine learning in automated text categorization}.
\newblock \bibinfo{journal}{\emph{Comput. Surveys}} \bibinfo{volume}{34},
  \bibinfo{number}{1} (\bibinfo{year}{2002}), \bibinfo{pages}{1--47}.
\newblock


\bibitem[\protect\citeauthoryear{Speicher, Heidari, Grgic{-}Hlaca, Gummadi,
  Singla, Weller, and Zafar}{Speicher et~al\mbox{.}}{2018}]%
        {kddSpeicherHGGSWZ18}
\bibfield{author}{\bibinfo{person}{Till Speicher}, \bibinfo{person}{Hoda
  Heidari}, \bibinfo{person}{Nina Grgic{-}Hlaca}, \bibinfo{person}{Krishna~P.
  Gummadi}, \bibinfo{person}{Adish Singla}, \bibinfo{person}{Adrian Weller},
  {and} \bibinfo{person}{Muhammad~Bilal Zafar}.}
  \bibinfo{year}{2018}\natexlab{}.
\newblock \showarticletitle{A unified approach to quantifying algorithmic
  unfairness: measuring individual {\&}group unfairness via inequality
  indices}. In \bibinfo{booktitle}{\emph{Proceedings of the 24th {ACM} {SIGKDD}
  International Conference on Knowledge Discovery {\&} Data Mining, {KDD} 2018,
  London, UK, August 19-23, 2018}}. \bibinfo{pages}{2239--2248}.
\newblock


\bibitem[\protect\citeauthoryear{Udeshi, Arora, and Chattopadhyay}{Udeshi
  et~al\mbox{.}}{2018}]%
        {UdeshiAC18}
\bibfield{author}{\bibinfo{person}{Sakshi Udeshi}, \bibinfo{person}{Pryanshu
  Arora}, {and} \bibinfo{person}{Sudipta Chattopadhyay}.}
  \bibinfo{year}{2018}\natexlab{}.
\newblock \showarticletitle{Automated directed fairness testing}. In
  \bibinfo{booktitle}{\emph{Proceedings of the 33rd {ACM/IEEE} International
  Conference on Automated Software Engineering, {ASE} 2018}}.
  \bibinfo{pages}{98--108}.
\newblock


\bibitem[\protect\citeauthoryear{Wick, Panda, and Tristan}{Wick
  et~al\mbox{.}}{2019}]%
        {nipsWickpT19}
\bibfield{author}{\bibinfo{person}{Michael~L. Wick},
  \bibinfo{person}{Swetasudha Panda}, {and} \bibinfo{person}{Jean{-}Baptiste
  Tristan}.} \bibinfo{year}{2019}\natexlab{}.
\newblock \showarticletitle{Unlocking fairness: a trade-off revisited}. In
  \bibinfo{booktitle}{\emph{Proceedings of the Annual Conference on Neural
  Information Processing Systems 2019, NeurIPS 2019}}.
  \bibinfo{pages}{8780--8789}.
\newblock


\bibitem[\protect\citeauthoryear{Yao and Shepperd}{Yao and Shepperd}{2020}]%
        {easeYaoS20}
\bibfield{author}{\bibinfo{person}{Jingxiu Yao} {and}
  \bibinfo{person}{Martin~J. Shepperd}.} \bibinfo{year}{2020}\natexlab{}.
\newblock \showarticletitle{Assessing software defection prediction
  performance: why using the Matthews correlation coefficient matters}. In
  \bibinfo{booktitle}{\emph{Proceedings of Evaluation and Assessment in
  Software Engineering, {EASE} 2020}}. \bibinfo{pages}{120--129}.
\newblock


\bibitem[\protect\citeauthoryear{Zafar, Valera, Gomez{-}Rodriguez, and
  Gummadi}{Zafar et~al\mbox{.}}{2017}]%
        {aistatsZafarVGG17}
\bibfield{author}{\bibinfo{person}{Muhammad~Bilal Zafar},
  \bibinfo{person}{Isabel Valera}, \bibinfo{person}{Manuel Gomez{-}Rodriguez},
  {and} \bibinfo{person}{Krishna~P. Gummadi}.} \bibinfo{year}{2017}\natexlab{}.
\newblock \showarticletitle{Fairness constraints: mechanisms for fair
  classification}. In \bibinfo{booktitle}{\emph{Proceedings of the 20th
  International Conference on Artificial Intelligence and Statistics, {AISTATS}
  2017}}. \bibinfo{pages}{962--970}.
\newblock


\bibitem[\protect\citeauthoryear{Zemel, Wu, Swersky, Pitassi, and Dwork}{Zemel
  et~al\mbox{.}}{2013}]%
        {LFRpaper}
\bibfield{author}{\bibinfo{person}{Richard~S. Zemel}, \bibinfo{person}{Yu Wu},
  \bibinfo{person}{Kevin Swersky}, \bibinfo{person}{Toniann Pitassi}, {and}
  \bibinfo{person}{Cynthia Dwork}.} \bibinfo{year}{2013}\natexlab{}.
\newblock \showarticletitle{Learning fair representations}. In
  \bibinfo{booktitle}{\emph{Proceedings of the 30th International Conference on
  Machine Learning, {ICML} 2013}}. \bibinfo{pages}{325--333}.
\newblock


\bibitem[\protect\citeauthoryear{Zhang, Lemoine, and Mitchell}{Zhang
  et~al\mbox{.}}{2018}]%
        {ADVpaper}
\bibfield{author}{\bibinfo{person}{Brian~Hu Zhang}, \bibinfo{person}{Blake
  Lemoine}, {and} \bibinfo{person}{Margaret Mitchell}.}
  \bibinfo{year}{2018}\natexlab{}.
\newblock \showarticletitle{Mitigating unwanted biases with adversarial
  learning}. In \bibinfo{booktitle}{\emph{Proceedings of the 2018 {AAAI/ACM}
  Conference on AI, Ethics, and Society, {AIES} 2018}}.
  \bibinfo{pages}{335--340}.
\newblock


\bibitem[\protect\citeauthoryear{Zhang and Harman}{Zhang and Harman}{2021}]%
        {icseZhangH21}
\bibfield{author}{\bibinfo{person}{Jie~M. Zhang} {and} \bibinfo{person}{Mark
  Harman}.} \bibinfo{year}{2021}\natexlab{}.
\newblock \showarticletitle{Ignorance and prejudice in software fairness}. In
  \bibinfo{booktitle}{\emph{Proceedings of the 43rd {IEEE/ACM} International
  Conference on Software Engineering, {ICSE} 2021}}.
  \bibinfo{pages}{1436--1447}.
\newblock


\bibitem[\protect\citeauthoryear{Zhang, Harman, Ma, and Liu}{Zhang
  et~al\mbox{.}}{2019}]%
        {jieMLsurvey}
\bibfield{author}{\bibinfo{person}{Jie~M. Zhang}, \bibinfo{person}{Mark
  Harman}, \bibinfo{person}{Lei Ma}, {and} \bibinfo{person}{Yang Liu}.}
  \bibinfo{year}{2019}\natexlab{}.
\newblock \showarticletitle{Machine learning testing: survey, landscapes and
  horizons}.
\newblock \bibinfo{journal}{\emph{IEEE Transactions on Software Engineering}}
  (\bibinfo{year}{2019}).
\newblock


\bibitem[\protect\citeauthoryear{Zhang, Zhang, and Zhang}{Zhang
  et~al\mbox{.}}{2021}]%
        {isstaZhangZZ21}
\bibfield{author}{\bibinfo{person}{Lingfeng Zhang}, \bibinfo{person}{Yueling
  Zhang}, {and} \bibinfo{person}{Min Zhang}.} \bibinfo{year}{2021}\natexlab{}.
\newblock \showarticletitle{Efficient white-box fairness testing through
  gradient search}. In \bibinfo{booktitle}{\emph{Proceedings of the 30th {ACM}
  {SIGSOFT} International Symposium on Software Testing and Analysis, ISSTA
  2021}}. \bibinfo{pages}{103--114}.
\newblock


\bibitem[\protect\citeauthoryear{Zhang and Sun}{Zhang and Sun}{2022}]%
        {mengdi22}
\bibfield{author}{\bibinfo{person}{Mengdi Zhang} {and} \bibinfo{person}{Jun
  Sun}.} \bibinfo{year}{2022}\natexlab{}.
\newblock \showarticletitle{Adaptive fairness improvement based on causality
  analysis}. In \bibinfo{booktitle}{\emph{Proceedings of the 30th {ACM} Joint
  European Software Engineering Conference and Symposium on the Foundations of
  Software Engineering, ESEC/FSE 2022}}.
\newblock


\bibitem[\protect\citeauthoryear{Zhang, Wang, Sun, Dong, Wang, Wang, Dong, and
  Dai}{Zhang et~al\mbox{.}}{2020}]%
        {icseZhangW0D0WDD20}
\bibfield{author}{\bibinfo{person}{Peixin Zhang}, \bibinfo{person}{Jingyi
  Wang}, \bibinfo{person}{Jun Sun}, \bibinfo{person}{Guoliang Dong},
  \bibinfo{person}{Xinyu Wang}, \bibinfo{person}{Xingen Wang},
  \bibinfo{person}{Jin~Song Dong}, {and} \bibinfo{person}{Ting Dai}.}
  \bibinfo{year}{2020}\natexlab{}.
\newblock \showarticletitle{White-box fairness testing through adversarial
  sampling}. In \bibinfo{booktitle}{\emph{Proceedings of the 42nd International
  Conference on Software Engineering, ICSE 2020}}. \bibinfo{pages}{949--960}.
\newblock


\bibitem[\protect\citeauthoryear{Zheng, Chen, Du, Zhang, Cheng, Ji, Wang, Yu,
  and Chen}{Zheng et~al\mbox{.}}{2022}]%
        {icseZhengCD0CJW0C22}
\bibfield{author}{\bibinfo{person}{Haibin Zheng}, \bibinfo{person}{Zhiqing
  Chen}, \bibinfo{person}{Tianyu Du}, \bibinfo{person}{Xuhong Zhang},
  \bibinfo{person}{Yao Cheng}, \bibinfo{person}{Shouling Ji},
  \bibinfo{person}{Jingyi Wang}, \bibinfo{person}{Yue Yu}, {and}
  \bibinfo{person}{Jinyin Chen}.} \bibinfo{year}{2022}\natexlab{}.
\newblock \showarticletitle{NeuronFair: Interpretable White-Box Fairness
  Testing through Biased Neuron Identification}. In
  \bibinfo{booktitle}{\emph{Proceedings of the 44th {IEEE/ACM} 44th
  International Conference on Software Engineering, {ICSE} 2022}}.
  \bibinfo{pages}{1519--1531}.
\newblock


\end{thebibliography}

\end{document}